\documentclass[aps,prx,reprint,amsmath,amssymb,longbibliography]{revtex4-2}
\usepackage[T1]{fontenc}
\usepackage{lmodern}
\usepackage{graphicx}
\usepackage[hidelinks]{hyperref}
\usepackage{tikz}
\newcommand{\Z}{\mathbb Z}
\newcommand{\Q}{\mathbb Q}
\newcommand{\C}{\mathbb C}
\newcommand{\cA}{\mathcal A}
\newtheorem{theorem}{Theorem}
\newtheorem{proposition}[theorem]{Proposition}
\newtheorem{corollary}[theorem]{Corollary}
\makeatletter
\AtBeginDocument{\immediate\write\@auxout{\string\citation{isingControl}}}
\makeatother

\begin{document}
\title{Galois conjugation, droplet growth, and the differential transcendence of the Ising model in nonzero field}

\author{Ran Luo$^2$}
\author{Jiahua Tian$^1$}

\affiliation{%
 $^1$School of Physics, East China Normal University, Shanghai, China, 200241
}
\affiliation{%
 $^2$School of Physics, Peking University, Beijing, China, 100871
}

\begin{abstract}
The square-lattice Ising model is exactly solvable in zero magnetic field, but the scope of exact solutions at nonzero field remains unresolved. We show that, at every fixed dimensionless field $H=h/(k_BT)\ne 0$, its free energy and magnetization obey no algebraic differential equation in the coupling $K=J/(k_BT)$. This excludes Onsager-type formulas and broad classes of special-function expressions. Previous work left open a countable set of fields with algebraic-integer fugacities. We eliminate these exceptions using Galois conjugation and the large-order growth of droplets of reversed spins. At conjugate fugacities outside the unit circle, nearly square droplets produce coefficient growth incompatible with an algebraic differential equation. We prove that cancellations cannot remove this growth in the full pressure series. The free-energy result also holds for the Lieb lattice and square lattices with equal-length chain decorations and uniform coupling and field. In every dimension $d\ge2$, we additionally resolve fields for which $e^{-2|H|}$ has a real algebraic conjugate greater than one.

\end{abstract}
\maketitle

\section{Introduction}
\label{sec:intro}
It is a long-standing question whether the free energy of the Ising model in a magnetic field can be written in closed form. The model in a field is the simplest system in which to study how integrable models relate to the generic, nonintegrable models that describe experiments~\cite{McCoyMaillard2012}. The expected natural boundary of the zero-field susceptibility~\cite{GuttmannEnting1996,Nickel1999,Nickel2000,OrrickNickelGuttmannPerk2001} and the zeros of the partition function in a field~\cite{Assis2017} suggest that a closed form is very hard to find, if it exists at all.

In a sharp form of the question we ask whether the free energy obeys an algebraic differential equation in the coupling, because the familiar forms of exact solutions all do~\cite{OrrickNickelGuttmannPerk2001,WuMcCoyTracyBarouch1976,JimboMiwa1980,McCoyMaillard2012,Guttmann2016}. We write $K=J/(k_BT)$ for the dimensionless coupling and $H=h/(k_BT)$ for the dimensionless field, and we vary $K$ at fixed $H$. A cooling path at fixed $J$ and $h$ instead keeps $H/K=h/J$ fixed. Closed forms built by arithmetic operations and composition from elliptic, hypergeometric, or Painlev\'e functions with fixed parameters, with arguments that are differentially algebraic in $K$, obey such equations as well~\cite{DLMF,WuMcCoyTracyBarouch1976,Manssour2024}.

It was recently shown that the free energy in a field obeys no algebraic differential equation in $K$, at every fixed $H$ outside a countable set and in every dimension $d\ge2$~\cite{LuoTian2026}. The obstruction comes from droplets of reversed spins. On the square lattice an $r\times r$ droplet contributes $u^{2r}z^{r^2}$ to the low-temperature series in $u=e^{-4K}$, where $z=e^{-2H}$ is the fugacity. At a rational fugacity $z=A/B$ these droplets force the coefficient of $u^{2r}$ to have a denominator divisible by $B^{r^2}$, which grows too fast for a solution of an algebraic differential equation~\cite{Polya1916,Popken1935,SibuyaSperber1981,KrattenthalerRivoal2025}.

The exceptional set consists of the fields whose fugacity is an algebraic integer, a root of a monic polynomial with integer coefficients. At such a fugacity the droplets produce no denominators, and the argument of~\cite{LuoTian2026} cannot be applied. Hence a closed form at a nonzero real field could occur only at such a fugacity. The set is countable, but it is dense in $(0,1)$ and contains, for instance, the golden-ratio fugacity $z=(\sqrt5-1)/2$. In this work we focus on these exceptions.

We resolve them completely on the square lattice and on decorated square lattices, and partially in every dimension. To do so we evaluate the low-temperature series at the Galois conjugates of the fugacity, the other roots of its minimal polynomial. Our results are the following.
\begin{enumerate}
\item On the square lattice, at every fixed nonzero real field, the free energy and the magnetization obey no algebraic differential equation on any interval of the coupling. Solutions of Onsager's type~\cite{LuoTian2026} and closed forms in elliptic, hypergeometric, or Painlev\'e functions with fixed parameters are therefore impossible at every nonzero field.
\item The free energies of the Lieb lattice and of every square lattice whose bonds are replaced by chains of equal length, with uniform coupling and field, obey no algebraic differential equation at any nonzero real field either. In zero field these free energies follow from Onsager's solution~\cite{Onsager1944}. In a nonzero field their effective fugacity varies with the temperature, so the square-lattice results do not apply to them, and we prove the result at every nonzero real field, including the fields whose fugacity is not an algebraic integer.
\item In every dimension $d\ge2$, we use a positivity argument to show that the formal low-temperature series at every real fugacity $q>1$, where the field favors minus spins, obeys no algebraic differential equation. Consequently, the free energy on $\Z^d$ obeys no algebraic differential equation at every algebraic fugacity with a real conjugate larger than one, a dense set of fields.
\end{enumerate}

We use the same square droplets as at rational fugacities, but we measure their weights with a different absolute value. Besides the ordinary absolute value, a number has an $\ell$-adic absolute value for each prime $\ell$, which is large when a high power of $\ell$ divides its denominator~\cite{Neukirch1999}. A nonzero algebraic number that is not a root of unity has either an $\ell$-adic absolute value larger than one or a Galois conjugate of modulus larger than one~\cite{Kronecker1857,Neukirch1999}. The first case gives the denominators of~\cite{LuoTian2026}. At an algebraic integer $\alpha\in(0,1)$ all $\ell$-adic absolute values are at most one, but some conjugate $\beta$ obeys $|\beta|>1$. A power series with coefficients in the field $\Q(\alpha)$ that obeys an algebraic differential equation obeys one with coefficients in that field~\cite{Mahler1976}, and Galois conjugation maps it to an equation for the conjugated series. By Maillet's theorem, the coefficients of every power-series solution of such an equation are bounded by a power of a factorial~\cite{Maillet1903,Malgrange1989}. The conjugated series has the coefficients of the low-temperature series evaluated at the fugacity $\beta$, where the weight $\beta^{|S|}$ of a droplet $S$ grows with its area. For real $\beta>1$ it is the formal low-temperature expansion of the plus phase in a field that favors minus spins, in which large droplets dominate, as in the nucleation of a metastable phase~\cite{Langer1967}. It is shown in~\cite{Isakov1984,FriedliPfister2004} that at low temperature large droplets make the $k$th derivative of the free energy with respect to the field grow like $(k!)^{d/(d-1)}$ at $H=0$, so that the free energy has no analytic continuation across $H=0$.

Prellberg and Owczarek~\cite{PrellbergOwczarek1999} showed that, for every fixed ($q>1$), near-square convex polygons determine the leading large-perimeter asymptotics of the sum over all square-lattice self-avoiding polygons weighted by ($q^{\mathrm{area}}$). This ensemble models inflated lattice vesicles~\cite{FisherGuttmannWhittington1991}. The same pattern is observed in~\cite{Meissner2023} for the low-temperature expansion of the Ising model in a field. We show that it holds exactly for the full free energy, in which droplets with holes and all Mayer clusters also contribute, and that all other droplets and clusters are negligible at every complex fugacity outside the unit circle, not only at real $q>1$, where the argument of~\cite{PrellbergOwczarek1999} uses the positivity of the weights. The generating function of the near-square droplets is a theta function times the fourth power of the partition generating function~\cite{PrellbergOwczarek1999}, and by Jacobi's triple product it has no zeros in the unit disk~\cite{DLMF}. Hence the near-square droplets cannot cancel at any conjugate, and the coefficients of the conjugated series grow like $|\beta|^{r^2}$, too fast for Maillet's bound.

Combining this argument with~\cite{LuoTian2026}, we find that the formal low-temperature series of the square lattice obeys no algebraic differential equation at any complex fugacity other than zero and the roots of unity. The only real roots of unity are $z=1$ and $z=-1$, the two fugacities at which the free energy is known~\cite{Onsager1944,LeeYang1952}. In Sec.~\ref{sec:review} we recall the argument of~\cite{LuoTian2026} and explain why it fails at algebraic integers. In Sec.~\ref{sec:proof} we prove our result at algebraic-integer fugacities on the square lattice. In Sec.~\ref{sec:higher} we treat real conjugates in every dimension and coupled layers, in Sec.~\ref{sec:decorated} we treat decorated lattices, and in Sec.~\ref{sec:conclusion} we discuss the physical meaning of our results and the open questions.

\section{Droplets and denominators}
\label{sec:review}
In this section we summarize the argument of~\cite{LuoTian2026} for fugacities that are not algebraic integers. We consider the Ising model on the square lattice $\Z^2$, with spins $\sigma_i=\pm1$ and Hamiltonian
\begin{equation}
	\mathcal H=-J\sum_{\langle ij\rangle}\sigma_i\sigma_j-h\sum_i\sigma_i,\qquad J>0,
	\label{eq:Hamiltonian}
\end{equation}
where $\langle ij\rangle$ runs over the nearest-neighbor bonds. Spin reversal maps $h$ to $-h$, so we take $h>0$. The dimensionless pressure $p(K,H)=\lim_{N\to\infty}N^{-2}\log Z_N$ of $N\times N$ boxes, which gives the free energy per site $f=-k_BT\,p$, has the low-temperature expansion~\cite{SykesEssamGaunt1965}
\begin{equation}
	p=2K+H+\Phi(u,z),\qquad \Phi(u,z)=\sum_{m\ge2}L_m(z)\,u^m,
	\label{eq:series}
\end{equation}
where $2K+H$ is the pressure of the fully aligned state. Relative to that state, a set $S$ of reversed spins with $|\partial S|$ boundary bonds has the Boltzmann weight $u^{m(S)}z^{|S|}$ with $m(S)=|\partial S|/2$~\cite{Enting1978}, so $u$ is the weight of two broken bonds and $z$ that of one reversed spin. A single reversed spin has $|\partial S|=4$, and every nonempty $S$ has $|\partial S|\ge4$, since its leftmost, rightmost, lowest, and highest sites each have a boundary bond pointing out of $S$. Hence $m(S)\ge2$, and the sum in~\eqref{eq:series} starts at $m=2$. For every $|z|<1$ the series~\eqref{eq:series} converges near $u=0$ to the physical $\Phi$, and its sum is analytic in $(u,z)$ jointly, which was shown with the Lee--Yang circle theorem~\cite{LeeYang1952,LuoTian2026}.

A function $\Phi(u)$ is differentially algebraic if it obeys
\begin{equation}
	\sum_{\mu,\nu}c_{\mu,\nu}\,u^{\mu}\prod_{j=0}^{s}\bigl(\partial_u^j\Phi\bigr)^{\nu_j}=0
	\label{eq:DA}
\end{equation}
for some order $s$, with finitely many nonnegative integer exponents $\mu$ and $\nu=(\nu_0,\ldots,\nu_s)$ and complex coefficients $c_{\mu,\nu}$, not all zero. Otherwise $\Phi$ is differentially transcendental~\cite{Rubel1989}. Differentially algebraic functions are closed under arithmetic operations and composition~\cite{Manssour2024}, so differential algebraicity in $u$ and in $K$ are equivalent. The pressure is real analytic in $(K,H)$ jointly for $K>0$ and $H\ne0$~\cite{Ott2020}, so at fixed $H\ne0$ an equation~\eqref{eq:DA} on a short interval of $K$ extends to all $K>0$, and then to the series~\eqref{eq:series} near $u=0$~\cite{LuoTian2026}. It therefore suffices to show that, at fixed $z$, the series~\eqref{eq:series} obeys no equation~\eqref{eq:DA}.

To describe the coefficients $L_m$, we call the connected components of $S$ its droplets. In an $a\times b$ box $\Lambda_{a,b}$ with plus boundary spins,
\begin{equation}
	P_{a,b}(u,z)=\sum_{S\subseteq\Lambda_{a,b}}u^{m(S)}z^{|S|}\in\Z[u,z],\quad P_{a,b}(0,z)=1,
	\label{eq:finiteP}
\end{equation}
is the partition function of a gas of droplets with hard-core exclusion, because the droplets of one configuration neither overlap nor touch~\cite{GruberKunz1971}. Its logarithm is the Mayer expansion~\cite{Mayer1937,Brydges1986,Bissacot2010}
\begin{equation}
	\log P_{a,b}=\sum_{k\ge1}\frac1{k!}\sum_{\gamma_1,\ldots,\gamma_k}\varphi^T(\gamma_1,\ldots,\gamma_k)\prod_{i=1}^ku^{m(\gamma_i)}z^{|\gamma_i|}.
	\label{eq:mayer}
\end{equation}
Two droplets are incompatible when they share a site or contain neighboring sites, and every droplet is incompatible with itself. The truncated function $\varphi^T$ is the sum of $(-1)^{|E(G)|}$ over the connected graphs $G$ on the labels $1,\ldots,k$ whose edges join incompatible droplets. A list of droplets with $\varphi^T\ne0$ is a Mayer cluster. Repeats are allowed, and a single droplet is a Mayer cluster with $\varphi^T=1$. Counting the translates of Mayer clusters in the box, we have~\cite{LuoTian2026}
\begin{equation}
	L_m(z)=\sum_{\mathcal C}\frac1{k!}\,\varphi^T(\gamma_1,\ldots,\gamma_k)\,z^{\sum_i|\gamma_i|},
	\label{eq:Lm}
\end{equation}
where $\mathcal C$ runs over the finitely many classes of Mayer clusters with $\sum_im(\gamma_i)=m$ under translations. Since $P_{a,b}$ has integer coefficients and $P_{a,b}(0,z)=1$, the logarithmic derivative $u\partial_u\log P_{a,b}$ has integer coefficients, so $m[u^m]\log P_{a,b}\in\Z[z]$. The union of the droplets of each class in~\eqref{eq:Lm} has a bounding box $w\times h$ with $w,h\le m$, which has $(a-w+1)(b-h+1)$ translates in $\Lambda_{a,b}$. For $a,b\ge m$ the coefficient $[u^m]\log P_{a,b}$ is therefore a polynomial of degree one in each of $a$ and $b$, with the coefficient $L_m$ of $ab$, and with $\Delta_af(a)=f(a+1)-f(a)$ we obtain~\cite{LuoTian2026}
\begin{equation}
	mL_m=\Delta_a\Delta_b\bigl(m[u^m]\log P_{a,b}\bigr)\in\Z[z].
	\label{eq:mLm}
\end{equation}

A droplet that occupies $w_1$ columns and $w_2$ rows has at least two boundary bonds in each of them, so $m(S)\ge w_1+w_2$ and
\begin{equation}
	|S|\le w_1w_2\le A_m:=\left\lfloor\frac{m^2}{4}\right\rfloor,\qquad m=m(S).
	\label{eq:area}
\end{equation}
Equality holds only for the $r\times r$ square at $m=2r$ and for the two $r\times(r+1)$ rectangles at $m=2r+1$. For $k\ge2$ droplets with half-perimeters $m_i\ge2$ and $\sum_im_i=m$, by moving half-perimeter from smaller to larger droplets we have~\cite{LuoTian2026}
\begin{equation}
	\sum_{i=1}^{k}A_{m_i}\le A_{m-2(k-1)}+k-1\le B_m:=A_m-(m-2).
	\label{eq:gap}
\end{equation}
Hence only single droplets contribute to $[z^n]L_m$ in the integer band $B_m<n\le A_m$, each with coefficient one per translation class. In particular $[z^{r^2}]L_{2r}=1$.

At a rational fugacity $z=A/B$ with coprime $A,B>0$ and $B>1$, let $\ell$ be a prime that divides $B$, let $v_\ell$ be the $\ell$-adic valuation, and put $a_\ell=v_\ell(B)$. The square term $z^{r^2}$ of $L_{2r}$ has $v_\ell=-a_\ell r^2$. Every other term of the integer band has an integer coefficient and $v_\ell\ge-a_\ell(r^2-1)$. The other terms are powers $z^n$ with $n\le B_{2r}=r^2-2r+2$, by~\eqref{eq:gap}, whose coefficients have denominators dividing $2r$, so they have $v_\ell\ge-a_\ell(r^2-2r+2)-v_\ell(2r)$. Since $a_\ell(2r-2)>v_\ell(2r)$ for $r\ge3$, the square term alone has the most negative valuation, and because $v_\ell(x+y)=\min\{v_\ell(x),v_\ell(y)\}$ whenever $v_\ell(x)\ne v_\ell(y)$,
\begin{equation}
	v_\ell\bigl(L_{2r}(A/B)\bigr)=-v_\ell(B)\,r^2,\qquad r\ge3.
	\label{eq:valuation}
\end{equation}
Hence the reduced denominator of $L_{2r}(A/B)$ is divisible by $B^{r^2}$. A power series with algebraic coefficients $g_n$ that obeys an equation~\eqref{eq:DA} has denominators $d_n$ with $\log d_n=O(n\log^2n)$~\cite{Polya1916,Popken1935,Mahler1976,KrattenthalerRivoal2025}, and at every prime its coefficients obey $v(g_n)\ge-C(n+1)$ with a constant $C$~\cite{SibuyaSperber1981}. The growth $r^2$ in~\eqref{eq:valuation} violates both bounds.

The same reasoning applies to every algebraic fugacity $\alpha$ that is not an algebraic integer. Some valuation $v$ of the number field $\Q(\alpha)$ that extends $v_\ell$ for a prime $\ell$ then has $v(\alpha)<0$~\cite{Neukirch1999}. It obeys the same rules as $v_\ell$, and the comparison that led to~\eqref{eq:valuation} gives $v\bigl(L_{2r}(\alpha)\bigr)=v(\alpha)\,r^2$ for large $r$, again violating both bounds~\cite{LuoTian2026}. At a transcendental fugacity we fix the finitely many monomials of a hypothetical equation~\eqref{eq:DA} and keep $z$ as a variable. Since $L_m\in\Q[z]$, the Taylor coefficients of each monomial are polynomials in $z$ with rational coefficients, and~\eqref{eq:DA} becomes an infinite linear system for the coefficients $c_{\mu,\nu}$. At a rational $z_0\in(0,1)$ its only solution is zero, so some square minor $D(z)\in\Q[z]$ of the system obeys $D(z_0)\ne0$. A nonzero polynomial with rational coefficients does not vanish at a transcendental number, so the system has only the zero solution at every transcendental $z$ as well~\cite{LuoTian2026}.

We cannot apply either argument to an algebraic integer $\alpha\in(0,1)$. Every valuation $v$ of $\Q(\alpha)$ that extends some $v_\ell$ obeys $v(\alpha)\ge0$~\cite{Neukirch1999}, so every power $\alpha^n$, including the square term $\alpha^{r^2}$, is an algebraic integer and contributes no denominator, unlike $(A/B)^{r^2}$. The only denominators of $L_{2r}(\alpha)$ come from the rational coefficients of $L_{2r}(z)$. Since $2rL_{2r}\in\Z[z]$, the number $2rL_{2r}(\alpha)$ is an algebraic integer, so
\begin{equation}
	v\bigl(L_{2r}(\alpha)\bigr)\ge-v(2r)\ge-\log_\ell(2r)
	\label{eq:smallden}
\end{equation}
for every such $v$, and the denominator of $L_{2r}(\alpha)$ divides $2r$. This growth is far below the bounds of~\cite{Polya1916,Popken1935,SibuyaSperber1981}, so these bounds give no contradiction. The minor $D(z)$ may vanish at an algebraic number, so the transcendental argument does not apply either. Nor does the ordinary absolute value of $\alpha$ help. Since $|\alpha|<1$, the square term $\alpha^{r^2}$, which is the highest power of $\alpha$ in $L_{2r}(\alpha)$, is its smallest monomial rather than its largest, and the series~\eqref{eq:series} converges. Therefore, at an algebraic-integer fugacity the square droplets leave no trace in the denominators or in the size of the coefficients $L_m(\alpha)$. The fields $H=-\frac12\log\alpha$ of these fugacities form a countable set that is dense on the positive axis, and in~\cite{LuoTian2026} they are left open.

\section{Algebraic-integer fugacities}
\label{sec:proof}
Our main result is the following.
\begin{theorem}
\label{thm:square}
Let $H\ne0$ be real. On the square lattice, the pressure $p(K,H)$ and the magnetization $M=\partial_Hp$ obey no algebraic differential equation in $K$ on any nonempty open interval of $K>0$.
\end{theorem}
At the fields whose fugacity $e^{-2|H|}$ is not an algebraic integer, this is shown in~\cite{LuoTian2026}. To prove Theorem~\ref{thm:square} at the remaining fields, we fix an algebraic integer $\alpha$ with $0<\alpha<1$ and show that the low-temperature series $\Phi(u,\alpha)$ of the square lattice obeys no equation~\eqref{eq:DA}. We first pass to a Galois conjugate of $\alpha$ outside the unit circle and reduce the problem to the growth of the conjugated coefficients, a step that does not depend on the lattice. We then count the near-square droplets that give the highest powers of $z$ in $L_{2r}$, and we bound all other droplets and Mayer clusters uniformly. With these bounds we show that the near-square droplets dominate the coefficients at the conjugate.

\subsection{Conjugate fugacities}
\label{sec:conjugate}
Let $P(x)\in\Q[x]$ be the monic minimal polynomial of $\alpha$ over $\Q$,
\begin{equation}
    P(x)=x^n+a_{n-1}x^{n-1}+\cdots+a_1x+a_0,
    \qquad a_j\in\Z,
    \label{eq:minpoly}
\end{equation}
where the integrality of its coefficients follows from $\alpha$ being an algebraic integer~\cite{Neukirch1999}, and $n\ge2$ because no integer lies in $(0,1)$. The smallest field that contains $\Q$ and $\alpha$ is
\begin{equation}
	F=\Q(\alpha)=\Bigl\{\sum_{k=0}^{n-1}c_k\alpha^k:\ c_k\in\Q\Bigr\},
	\label{eq:field}
\end{equation}
and each of its elements has a unique representation of this form. A field embedding of $F$ is a map $\sigma:F\to\C$ with
\begin{equation}
	\sigma(x+y)=\sigma(x)+\sigma(y),\quad \sigma(xy)=\sigma(x)\,\sigma(y),\quad \sigma(1)=1.
	\label{eq:embedding}
\end{equation}
It fixes every rational number, so $\sigma\bigl(\sum_kc_k\alpha^k\bigr)=\sum_kc_k\,\sigma(\alpha)^k$, and it is determined by $\sigma(\alpha)$. Applying $\sigma$ to $P(\alpha)=0$ gives $P(\sigma(\alpha))=0$. Conversely, for every root $\gamma$ of $P$ the map $\sum_kc_k\alpha^k\mapsto\sum_kc_k\gamma^k$ is a field embedding, since it is well-defined by the uniqueness in~\eqref{eq:field} and multiplicative because products in~\eqref{eq:field} are reduced modulo $P$ and $P(\gamma)=0$. The roots of $P$ are distinct because $P$ is irreducible over $\Q$. Hence $F$ has exactly $n$ embeddings $\sigma_1,\ldots,\sigma_n$, their values $\sigma_i(\alpha)$ are the $n$ roots of $P$, and
\begin{equation}
	P(x)=\prod_{i=1}^{n}\bigl(x-\sigma_i(\alpha)\bigr).
	\label{eq:factor}
\end{equation}
These roots are the Galois conjugates of $\alpha$. We let $\sigma_1(\alpha)=\alpha$, and by Vieta's theorem
\begin{equation}
	\prod_{i=1}^{n}\sigma_i(\alpha)=(-1)^n a_0.
	\label{eq:norm}
\end{equation}
Since $P$ is irreducible and $n\ge2$, $x$ does not divide $P$, so $a_0$ is a nonzero integer and $|a_0|\ge1$. If $|\sigma_i(\alpha)|\le1$ for all $i\ge2$, then
\begin{equation}
	1\le\prod_{i=1}^{n}\bigl|\sigma_i(\alpha)\bigr|=\alpha\prod_{i=2}^{n}\bigl|\sigma_i(\alpha)\bigr|\le\alpha<1,
	\label{eq:normineq}
\end{equation}
which is impossible. Hence some embedding $\sigma$ obeys
\begin{equation}
	\beta=\sigma(\alpha),\qquad |\beta|>1.
	\label{eq:conjugate}
\end{equation}

Suppose that $\Phi(u,\alpha)$ obeys an equation~\eqref{eq:DA}. Its Taylor coefficients $L_m(\alpha)$ lie in $F$, so by statement~(6) of Sec.~25 of~\cite{Mahler1976} it also obeys an equation~\eqref{eq:DA} whose coefficients $c_{\mu,\nu}$ lie in $F$, see also Proposition~8 of Chap.~II of~\cite{Kolchin1973} and its Corollary. We fix $\sigma$ as in~\eqref{eq:conjugate} and let it act on formal power series with coefficients in $F$ coefficientwise,
\begin{equation}
	\sigma\Bigl(\sum_{m\ge0}b_m\,u^m\Bigr)=\sum_{m\ge0}\sigma(b_m)\,u^m,\qquad b_m\in F.
	\label{eq:sigmaseries}
\end{equation}
By~\eqref{eq:embedding}, this map preserves sums and products of series, because each coefficient of a product is a finite sum of products of coefficients. It also commutes with $\partial_u$, because $\sigma(m\,b_m)=m\,\sigma(b_m)$. Since $L_m\in\Q[z]$, we have $\sigma\bigl(L_m(\alpha)\bigr)=L_m\bigl(\sigma(\alpha)\bigr)=L_m(\beta)$, so~\eqref{eq:sigmaseries} maps $\Phi(u,\alpha)=\sum_{m\ge2}L_m(\alpha)\,u^m$ to the formal series
\begin{equation}
	\Phi_\beta(u)=\sum_{m\ge2}L_m(\beta)\,u^m.
	\label{eq:conjseries}
\end{equation}
Applying $\sigma$ to~\eqref{eq:DA} therefore gives
\begin{equation}
	\sum_{\mu,\nu}\sigma(c_{\mu,\nu})\,u^{\mu}\prod_{j=0}^{s}\bigl(\partial_u^j\Phi_\beta\bigr)^{\nu_j}=0.
	\label{eq:conjDA}
\end{equation}
A field embedding is injective, so the coefficients $\sigma(c_{\mu,\nu})$ are not all zero, and $\Phi_\beta$ obeys a nonzero algebraic differential equation.

We now bound the coefficients $L_m(\beta)$ with Maillet's theorem. Let $f(u)=\sum_{m\ge0}f_m\,u^m$ be a formal power series with $f_0\ne0$ that obeys a nonzero algebraic differential equation. By the theorem on p.~488 of~\cite{Maillet1903}, there are constants $C_0,B>0$ and $s_0\ge0$ such that
\begin{equation}
	|f_m|\le C_0\,B^m\,(m!)^{s_0},\qquad m\ge0,
	\label{eq:mailletthm}
\end{equation}
whether or not the series converges~\cite{Malgrange1989}, see also Sec.~1 of~\cite{GontsovGoryuchkina2018}. We take $f=1+\Phi_\beta$, so that $f_0=1$, $f_1=0$, and $f_m=L_m(\beta)$ for $m\ge2$. Substituting $\Phi_\beta=f-1$ and $\partial_u^j\Phi_\beta=\partial_u^jf$ for $j\ge1$ into~\eqref{eq:conjDA} gives
\begin{equation}
	\sum_{\mu,\nu}\sigma(c_{\mu,\nu})\,u^{\mu}\,(f-1)^{\nu_0}\prod_{j=1}^{s}\bigl(\partial_u^jf\bigr)^{\nu_j}=0.
	\label{eq:fDA}
\end{equation}
This equation is nonzero, because replacing $y_0$ by $y_0-1$ in the nonzero polynomial $\sum_{\mu,\nu}\sigma(c_{\mu,\nu})\,u^{\mu}\prod_{j=0}^{s}y_j^{\nu_j}$ gives a nonzero polynomial. Since $m!\le m^m$, the bound~\eqref{eq:mailletthm} gives, for every $m\ge2$ with $L_m(\beta)\ne0$,
\begin{equation}
	\log|L_m(\beta)|\le\log C_0+m\log B+s_0\,m\log m,
	\label{eq:mailletlog}
\end{equation}
and therefore
\begin{equation}
	\log^+|L_m(\beta)|=O(m\log m),
	\label{eq:maillet}
\end{equation}
with $\log^+x=\max\{\log x,0\}$. For all conjugates at once, the bound~\eqref{eq:maillet} is Theorem~17 of~\cite{Mahler1976}, see also~\cite{Popken1935}.

By~\eqref{eq:Lm}, $L_{2r}$ is a polynomial in $z$ without constant term. Every Mayer cluster at $m=2r$ has total area at most $A_{2r}=r^2$, by~\eqref{eq:area} for single droplets and by~\eqref{eq:gap} for clusters of several droplets. With the deficit $k=r^2-n$ below the area of the $r\times r$ square,
\begin{equation}
\begin{split}
	L_{2r}(z)&=\sum_{n=1}^{r^2}\bigl([z^n]L_{2r}\bigr)z^n\\
	&=z^{r^2}\sum_{k=0}^{r^2-1}\bigl([z^{r^2-k}]L_{2r}\bigr)z^{-k}.
\end{split}
	\label{eq:Lreverse}
\end{equation}
We therefore write $L_{2r}(\beta)=\beta^{r^2}F_r(1/\beta)$ with the polynomial
\begin{equation}
	F_r(t)=t^{r^2}L_{2r}(1/t)=\sum_{k\ge0}\bigl([z^{r^2-k}]L_{2r}\bigr)\,t^k,
	\label{eq:Fr}
\end{equation}
whose constant term $[z^{r^2}]L_{2r}=1$ is the weight of the square. Since $|1/\beta|<1$, it suffices to show that $F_r(1/\beta)$ tends to a nonzero limit as $r\to\infty$, because then $\log|L_{2r}(\beta)|=r^2\log|\beta|+O(1)$ violates~\eqref{eq:maillet}.

At a rational fugacity $z=A/B$, the $\ell$-adic valuation of $L_{2r}(A/B)$ for a prime $\ell$ that divides $B$ is fixed by the square term alone, as in~\eqref{eq:valuation}, because $v_\ell(x+y)=\min\{v_\ell(x),v_\ell(y)\}$ whenever $v_\ell(x)\ne v_\ell(y)$. The ordinary absolute value has no such property, since $|x+y|$ may be much smaller than $\max\{|x|,|y|\}$, and at $\beta$ the square term does not decide the size of $L_{2r}(\beta)$ on its own. The number of droplets with area $r^2-k$ grows with $k$, so the terms with $k\ge1$ can exceed the square term in modulus. The coefficients carry signs and $\beta$ may be negative or complex, so these terms may cancel. We therefore need the whole sum~\eqref{eq:Fr}. From here on we use only $|\beta|>1$, and not that $\beta$ is a conjugate of $\alpha$, and we prove the following asymptotics.
\begin{proposition}
\label{prop:asym}
For every complex $b$ with $|b|>1$, as $r\to\infty$,
\begin{gather}
	L_{2r}(b)=b^{r^2}\bigl[\cA(1/b)+o(1)\bigr],\nonumber\\
	bL_{2r}'(b)=r^2b^{r^2}\bigl[\cA(1/b)+o(1)\bigr],
	\label{eq:asym}
\end{gather}
where $\cA(t)=\sum_{j\in\Z}t^{j^2}\big/\prod_{n\ge1}(1-t^n)^4$ has no zeros in $|t|<1$.
\end{proposition}
To this end, we first count the near-square droplets.

\subsection{Near-square droplets}
\label{sec:nearsquare}
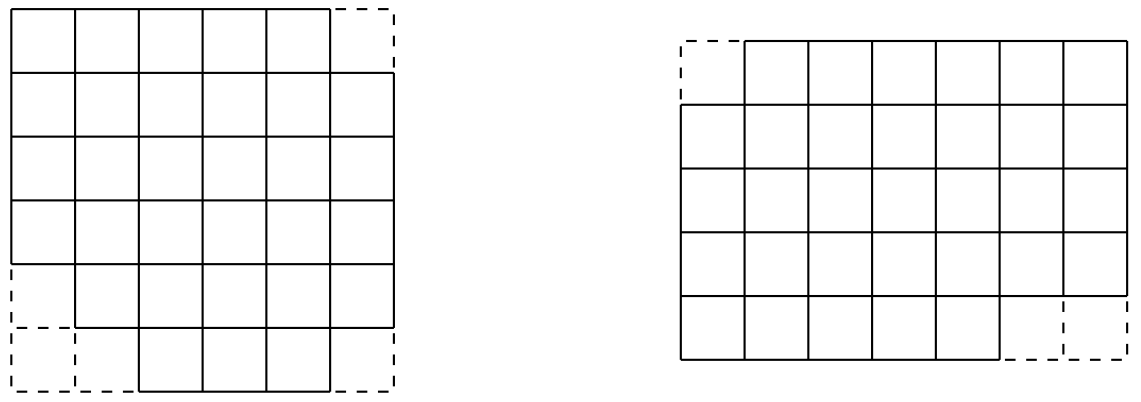
\begin{figure}[t]
\centering
\setlength{\unitlength}{1pt}
\begin{picture}(240,110)
 \put(14,24){%
  \begin{picture}(72,72)
   \put(24,0){\line(1,0){36}}
   \put(12,12){\line(1,0){60}}
   \put(0,24){\line(1,0){72}}
   \put(0,36){\line(1,0){72}}
   \put(0,48){\line(1,0){72}}
   \put(0,60){\line(1,0){72}}
   \put(0,72){\line(1,0){60}}
   \put(0,24){\line(0,1){48}}
   \put(12,12){\line(0,1){60}}
   \put(24,0){\line(0,1){72}}
   \put(36,0){\line(0,1){72}}
   \put(48,0){\line(0,1){72}}
   \put(60,0){\line(0,1){72}}
   \put(72,12){\line(0,1){48}}
   \multiput(1,0)(4,0){6}{\line(1,0){2}}
   \multiput(61,0)(4,0){3}{\line(1,0){2}}
   \multiput(1,12)(4,0){3}{\line(1,0){2}}
   \multiput(61,72)(4,0){3}{\line(1,0){2}}
   \multiput(0,1)(0,4){6}{\line(0,1){2}}
   \multiput(12,1)(0,4){3}{\line(0,1){2}}
   \multiput(72,1)(0,4){3}{\line(0,1){2}}
   \multiput(72,61)(0,4){3}{\line(0,1){2}}
  \end{picture}}
 \put(0,4){\makebox(100,12){\small (a) $6\times6$, $N=5$}}
 \put(140,30){%
  \begin{picture}(84,60)
   \put(0,0){\line(1,0){60}}
   \put(0,12){\line(1,0){84}}
   \put(0,24){\line(1,0){84}}
   \put(0,36){\line(1,0){84}}
   \put(0,48){\line(1,0){84}}
   \put(12,60){\line(1,0){72}}
   \put(0,0){\line(0,1){48}}
   \put(12,0){\line(0,1){60}}
   \put(24,0){\line(0,1){60}}
   \put(36,0){\line(0,1){60}}
   \put(48,0){\line(0,1){60}}
   \put(60,0){\line(0,1){60}}
   \put(72,12){\line(0,1){48}}
   \put(84,12){\line(0,1){48}}
   \multiput(61,0)(4,0){6}{\line(1,0){2}}
   \multiput(1,60)(4,0){3}{\line(1,0){2}}
   \multiput(0,49)(0,4){3}{\line(0,1){2}}
   \multiput(72,1)(0,4){3}{\line(0,1){2}}
   \multiput(84,1)(0,4){3}{\line(0,1){2}}
  \end{picture}}
 \put(132,4){\makebox(100,12){\small (b) $7\times5$, $N=3$}}
\end{picture}
\caption{Near-square droplets at half-perimeter $m=2r=12$. The missing cells are drawn with dashed lines. (a) The $6\times6$ square, $j=0$, after removing Young diagrams of sizes $3$, $1$, and $1$ at three corners, so that $k=N=5$. (b) The $7\times5$ rectangle, $j=1$, after removing diagrams of sizes $2$ and $1$, so that $k=N+j^2=4$. Every row and every column remains an interval, so both droplets keep $2m=24$ boundary bonds, and their areas are $r^2-k=31$ and $32$.}
\label{fig:corners}
\end{figure}

The droplets that we count are obtained from boxes of shape $(r+j)\times(r-j)$ by removing a Young diagram at each of the four corners, as in Fig.~\ref{fig:corners}. We encode them by ordered quadruples of partitions, one for each corner, and we first count these quadruples. Let $p(n)$ be the number of partitions of the integer $n\ge0$, with $p(n)=0$ for $n<0$, so that $\sum_{n\ge0}p(n)\,t^n=\prod_{n\ge1}(1-t^n)^{-1}$ by Euler's identity, see Eqs.~(27.14.2) and~(27.14.3) of~\cite{DLMF}. The number of ordered quadruples of partitions with total size $n$ is
\begin{equation}
	p_4(n)=\sum_{n_1+n_2+n_3+n_4=n}p(n_1)\,p(n_2)\,p(n_3)\,p(n_4),
	\label{eq:p4}
\end{equation}
and the Cauchy product of four copies of Euler's identity gives
\begin{equation}
	\sum_{n\ge0}p_4(n)\,t^n=\prod_{n\ge1}\frac{1}{(1-t^n)^4}.
	\label{eq:p4gen}
\end{equation}
The amplitude in Proposition~\ref{prop:asym} is
\begin{equation}
	\cA(t)=\frac{\sum_{j\in\Z}t^{j^2}}{\prod_{n\ge1}(1-t^n)^4}.
	\label{eq:amplitude}
\end{equation}
Multiplying the theta series $\sum_{j\in\Z}t^{j^2}$ by~\eqref{eq:p4gen} and collecting the terms with $j^2+N=k$ gives
\begin{equation}
\begin{split}
	\cA(t)&=\sum_{j\in\Z}\sum_{N\ge0}p_4(N)\,t^{j^2+N}\\
	&=\sum_{k\ge0}\Bigl(\sum_{j\in\Z}p_4(k-j^2)\Bigr)t^k.
\end{split}
	\label{eq:Agen}
\end{equation}
We show that for every $r\ge1$ and $0\le k<r$,
\begin{equation}
	[z^{r^2-k}]L_{2r}=\sum_{j\in\Z}p_4(k-j^2),
	\label{eq:stable}
\end{equation}
that is, by~\eqref{eq:Fr} and~\eqref{eq:Agen},
\begin{equation}
	F_r(t)-\cA(t)\in t^r\,\Q[[t]].
	\label{eq:Fstable}
\end{equation}
Since the coefficients of $F_r$ carry signs and $\beta$ may be negative or complex, estimates of the terms of $F_r(1/\beta)$ one by one bound it only from above and do not keep it away from zero. By~\eqref{eq:Fstable}, the first $r$ coefficients of $F_r$ do not depend on $r$ and are those of $\cA$. Therefore, if $\sum_{k\ge0}\bigl|[t^k]F_r\bigr|R^k$ is bounded uniformly in $r$ for some $R$ with $|\beta|^{-1}<R<1$, then $F_r(1/\beta)\to\cA(1/\beta)$, because $F_r-\cA$ starts at $t^r$ and so $|F_r(1/\beta)-\cA(1/\beta)|$ decays like $(R|\beta|)^{-r}$. This is the nonzero limit required in Sec.~\ref{sec:conjugate}, provided that $\cA(1/\beta)\ne0$.

To prove~\eqref{eq:stable}, let $\gamma_1,\gamma_2,\ldots$ be the droplets of a Mayer cluster with at least two droplets and $\sum_im(\gamma_i)=2r$. By~\eqref{eq:area} and~\eqref{eq:gap} at $m=2r$, where $A_{2r}=r^2$,
\begin{equation}
	\sum_i|\gamma_i|\le\sum_iA_{m(\gamma_i)}\le B_{2r}=A_{2r}-(2r-2)=r^2-2r+2.
	\label{eq:Bband}
\end{equation}
Here $k$ is the deficit of the area below $r^2$ in~\eqref{eq:stable}, with $0\le k<r$, not to be confused with the number of droplets in~\eqref{eq:gap}. For $r\ge2$,~\eqref{eq:Bband} gives $\sum_i|\gamma_i|\le r^2-r<r^2-k$. At $r=1$ every Mayer cluster is a single droplet, since two droplets need $m\ge4$. Hence, for $k<r$, the exponent $r^2-k$ lies in the integer band $B_{2r}<n\le A_{2r}$, where only single droplets contribute, each with coefficient one per translation class, and
\begin{equation}
	[z^{r^2-k}]L_{2r}=\#\Bigl(\bigl\{S:\ m(S)=2r,\ |S|=r^2-k\bigr\}\big/\Z^2\Bigr),
	\label{eq:count}
\end{equation}
the number of such droplets up to translation.

Let $w\times h$ be the bounding box of such a droplet $S$, so that $w+h\le m(S)=2r$ by~\eqref{eq:area}. If $w+h\le2r-1$, then
\begin{equation}
	|S|\le wh\le A_{2r-1}=r^2-r<r^2-k,
\end{equation}
which is impossible. Hence $w+h=2r$. Each of the $h$ rows contains at least two horizontal boundary bonds, and each of the $w$ columns at least two vertical ones, so
\begin{equation}
	4r=|\partial S|\ge2h+2w=4r,
	\label{eq:twobonds}
\end{equation}
and every row and every column contains exactly two boundary bonds. Every row and every column of $S$ is therefore an interval, and $S$ has no holes. We write $w=r+j$ and $h=r-j$ with $j\in\Z$. The box has area $wh=r^2-j^2$, so $S$ misses
\begin{equation}
	N=wh-|S|=k-j^2
	\label{eq:N}
\end{equation}
of its cells, and since $k<r$ and $j^2\ge|j|$,
\begin{equation}
	0\le N<r-j^2\le r-|j|=\min(w,h).
	\label{eq:Nbound}
\end{equation}

We place the box at $\{1,\ldots,w\}\times\{1,\ldots,h\}$ and write row $y$ of $S$ as the interval $\{a_y,\ldots,b_y\}$. By~\eqref{eq:Nbound}, at least one row $y_0$ and one column $x_0$ of the box are complete. Let $(x,y)$ be a missing cell with $x<x_0$ and $y<y_0$. Column $x$ is an interval that contains $(x,y_0)$, so $(x,y')$ is missing for every $y'\le y$. Row $y$ is an interval that contains $(x_0,y)$, so $(x',y)$ is missing for every $x'\le x$. Hence the numbers of missing cells
\begin{equation}
	\lambda_y=a_y-1,\qquad 1\le y<y_0,
	\label{eq:corner}
\end{equation}
at the left ends of the rows below $y_0$ obey $\lambda_1\ge\lambda_2\ge\cdots\ge\lambda_{y_0-1}\ge0$. They form a partition $\lambda^{(1)}$, whose Young diagram is attached to the corner $(1,1)$ of the box. The same argument at the other three corners gives partitions $\lambda^{(2)}$, $\lambda^{(3)}$, and $\lambda^{(4)}$ (Fig.~\ref{fig:corners}). Since row $y_0$ and column $x_0$ are complete, every missing cell lies in one of the four corner regions, and
\begin{equation}
	|\lambda^{(1)}|+|\lambda^{(2)}|+|\lambda^{(3)}|+|\lambda^{(4)}|=N.
	\label{eq:cornersum}
\end{equation}
No cell is missing between two complete rows or between two complete columns, so the four partitions do not depend on which complete row and column we choose.

Conversely, let $\lambda^{(1)},\ldots,\lambda^{(4)}$ be partitions that obey~\eqref{eq:cornersum} with $N<\min(w,h)$. A partition of size $n$ has at most $n$ parts, each at most $n$, so its diagram fits into an $n\times n$ square. Any two of the four diagrams therefore have total width at most $N<w$ and total height at most $N<h$. Hence the diagrams at the two lower corners share no column, those at the two left corners share no row, and the same holds at the upper and at the right corners. Each row and each column of the box loses only an initial and a final segment, of total length at most $N$. Every row and every column thus remains a nonempty interval, and some column remains complete and meets every row. The result is therefore a droplet with bounding box $w\times h$ and $|\partial S|=2w+2h=4r$, that is, $m(S)=2r$. Hence, by~\eqref{eq:p4}, each box $w\times h$ carries exactly
\begin{equation}
	\#\Bigl\{\bigl(\lambda^{(1)},\ldots,\lambda^{(4)}\bigr):\ \sum_{i=1}^{4}|\lambda^{(i)}|=N\Bigr\}=p_4(N)
	\label{eq:boxcount}
\end{equation}
droplets with $|S|=r^2-k$. Summing~\eqref{eq:boxcount} over $j\in\Z$ with $N=k-j^2$, and counting $j$ and $-j$ separately because rotated droplets lie in different translation classes, we obtain~\eqref{eq:stable} from~\eqref{eq:count}.

The function $\cA$ is the generating function of the near-square droplets. Indeed, in~\eqref{eq:Agen}, the theta series $\sum_jt^{j^2}$ records the shape $(r+j)\times(r-j)$ of the box, and each factor $\prod_n(1-t^n)^{-1}$ records the Young diagram at one corner. The same corner construction describes the convex polygons, and it gives the amplitude in Eqs.~(1.6)--(1.9) and~(4.1) of~\cite{PrellbergOwczarek1999}. For them, the agreement of the leading $\lfloor n/2\rfloor$ coefficients at half-perimeter $n$ with this amplitude is shown in the proof of Lemma~3.6 of~\cite{PrellbergOwczarek1999}. By Corollary~2.3 and Eq.~(2.8) of the same reference, the self-avoiding polygons $P$ of perimeter $4r$, counted up to translation and weighted by $q^{{\rm area}(P)}$ with $q>1$, obey
\begin{equation}
	\sum_{|\partial P|=4r}q^{{\rm area}(P)}\sim q^{r^2}\cA(1/q),\qquad r\to\infty.
	\label{eq:POasym}
\end{equation}
With~\eqref{eq:stable} we carry this to the free energy, where droplets with holes and Mayer clusters of several droplets also contribute, and by~\eqref{eq:Bband} and~\eqref{eq:twobonds} they do not reach the top $r$ coefficients. The first $r$ coefficients of~\eqref{eq:stable} are observed in the expansion of $e^\Phi$ in Eqs.~(18) and~(23) of~\cite{Meissner2023}. Each term $\Phi^l/l!$ with $l\ge2$ of $e^\Phi$ is a sum of products of $l$ coefficients $L_{m_i}$ with $\sum_im_i=2r$, whose degrees in $z$ add up to at most $B_{2r}$ by~\eqref{eq:gap}. Hence
\begin{equation}
	[u^{2r}z^n]\,e^{\Phi}=[u^{2r}z^n]\,\Phi,\qquad n>B_{2r},
	\label{eq:expPhi}
\end{equation}
and with~\eqref{eq:stable} we establish this observation.

\subsection{Uniform bounds}
\label{sec:bounds}
The stable coefficients determine the near-square contribution, but we must still control the sum of all larger deficits. At fixed total perimeter, holes and additional cluster members reduce the maximum available area. We show that, for every fixed $q>1$, the resulting loss of weight outweighs their shape and placement entropy.

Let $D_m(q)$ sum all single-droplet weights, let $Z_m(q)$ restrict this sum to droplets without holes, and let $E_m(q)$ sum the absolute weights of Mayer clusters with at least two members:
\begin{equation}
\begin{split}
 D_m(q)&=\sum_{S:\,m(S)=m}q^{|S|},\\
 E_m(q)&=\sum_{\substack{\mathcal C=(\gamma_1,\ldots,\gamma_k)\\
 k\ge2,\ \sum_i m(\gamma_i)=m}}
 \frac{|\varphi^T(\mathcal C)|}{k!}\,q^{\sum_i|\gamma_i|}.
\end{split}
\label{eq:DE}
\end{equation}
All sums are modulo overall translation. Our target is
\begin{equation}
 D_m(q)+E_m(q)\le C''_q q^{A_m},
 \qquad A_m=\lfloor m^2/4\rfloor,\quad m\ge2.
 \label{eq:keybound}
\end{equation}

\textit{Droplets without holes.}
The boundaries of simply-connected droplets are simple closed paths on the dual lattice, or self-avoiding polygons. The polygon theorem of Prellberg and Owczarek~\cite{PrellbergOwczarek1999}, Theorems~2.1--2.2 and Corollary~2.3, gives
\begin{equation}
 Z_m(q)\le C_q q^{A_m}\qquad(m\ge2)
 \label{eq:polygons}
\end{equation}
for every fixed $q>1$. This bound already includes all shapes without holes, however far from a square.

\textit{Droplets with holes.}
Holes are the finite nearest-neighbor components of the complement. If their total half-perimeter is $h$, filling them produces a droplet $\bar S$ without holes, with
\[
 m(\bar S)=m-h,\qquad |S|\le|\bar S|.
\]
Thus internal boundaries leave less perimeter for enclosing bulk area. Since each hole has half-perimeter at least two, there are at most $h/2$ holes. For fixed $\bar S$, their arrangements number at most
\begin{equation}
\underbrace{2^h}_{\substack{\text{compositions}\\\text{of holes}}}
\cdot
\underbrace{4^{2h}}_{\substack{\text{different directions}\\\text{of each walk}}}
\cdot
\underbrace{(m^2)^{h/2}}_{\text{displacements}}
=(32m)^h.
\end{equation}
Choosing one reference cell per hole gives the position bound, since $|\bar S|\le m^2$. Allowing intersecting or otherwise invalid arrangements only enlarges the count. Using $q^{|S|}\le q^{|\bar S|}$ and $A_m-A_{m-h}\ge mh/4-1$, we obtain, for sufficiently large $m$,
\begin{equation}
\begin{split}
 q^{-A_m}(D_m-Z_m)
 &\le C_q\sum_{h=2}^{m-2}(32m)^h q^{A_{m-h}-A_m}\\
 &\le qC_q\sum_{h\ge2}\bigl(32m q^{-m/4}\bigr)^h
 \longrightarrow0.
\end{split}
\label{eq:holes}
\end{equation}
The polynomial factor $32m$ loses to the exponential penalty $q^{-m/4}$. Absorbing finitely many small $m$ into the constant gives
\begin{equation}
 D_m(q)\le C'_q q^{A_m}\qquad(m\ge2).
 \label{eq:Cprime}
\end{equation}

\textit{Mayer clusters.}
These terms arise in $\log P$, so their members may overlap or repeat; their area exponent is the sum of the member areas, counted with multiplicity. In the incompatibility graph, each vertex represents one labelled droplet and an edge records overlap or nearest-neighbor contact. The Penrose tree bound controls $|\varphi^T|$ by the number of spanning trees of this graph~\cite{Penrose1967,FernandezProcacci2007}.

Fix one such tree and the position of its root. Each child must be incompatible with its parent: choosing one site in each gives five possible offsets, hence at most $5|\gamma_i||\gamma_j|\le5m^4$ relative positions. There are $k-1$ tree edges and $k^{k-2}$ labelled trees. Summing over shapes therefore gives
\begin{equation}
 E_m(q)\le
 \sum_{k=2}^{\lfloor m/2\rfloor}
 \frac{k^{k-2}}{k!}(5m^4)^{k-1}
 \sum_{\substack{m_1+\cdots+m_k=m\\m_i\ge2}}
 \prod_{i=1}^kD_{m_i}(q).
 \label{eq:Ebound}
\end{equation}
There are at most $m^{k-1}$ perimeter allocations. More importantly, the area bound~\eqref{eq:gap} gives
\[
 A_m-\sum_iA_{m_i}\ge(k-1)(m-k)\ge\frac{m(k-1)}2.
\]
Relative to one maximal droplet, each additional member thus costs at least a factor $q^{-m/2}$. Combining this loss with~\eqref{eq:Cprime} and $k^{k-2}/k!\le e^k$, we find, for sufficiently large $m$,
\begin{equation}
 q^{-A_m}E_m(q)
 \le eC'_q\sum_{k\ge2}
 \bigl(5eC'_q m^5q^{-m/2}\bigr)^{k-1}
 \longrightarrow0.
 \label{eq:clusters}
\end{equation}
Here too, the exponential area penalty dominates the polynomial counting factor. Together with~\eqref{eq:Cprime}, this proves~\eqref{eq:keybound}, after enlarging $C''_q$ to cover finitely many small $m$.

Finally, for $0<R<1$, set $q=1/R$ and define $\|G\|_R=\sum_j|g_j|R^j$ for $G(t)=\sum_jg_jt^j$. Taking absolute values before summing the cluster contributions gives
\begin{equation}
\begin{split}
 \|F_r\|_R
 &=q^{-r^2}\sum_n\bigl|[z^n]L_{2r}(z)\bigr|q^n\\
 &\le q^{-r^2}\bigl[D_{2r}(q)+E_{2r}(q)\bigr]
 \le C''_q.
\end{split}
\label{eq:normbound}
\end{equation}
Thus
\begin{equation}
 B_R:=\sup_{r\ge1}\|F_r\|_R<\infty,\qquad 0<R<1.
 \label{eq:BR}
\end{equation}
Because this is a bound on absolute coefficients, it controls complex phases as well as positive real weights.

\textit{Convergence and the amplitude.}
After normalization by the square weight, the power of $t$
measures the area deficit. Equation~\eqref{eq:Fstable} identifies
the coefficients at all deficits below $r$, while the uniform
bound~\eqref{eq:BR} controls the remaining contributions.
For fixed $0<\rho<R<1$, write
$F_r(t)-\cA(t)=\sum_{j\ge r}g_{r,j}t^j$.
Comparison of the weights at the two radii gives, for $|t|\le\rho$,
\begin{equation}
    \bigl|F_r(t)-\cA(t)\bigr|
    \le \sum_{j\ge r}|g_{r,j}|\,\rho^j
    \le \Bigl(\frac{\rho}{R}\Bigr)^r
        \|F_r-\cA\|_R.
    \label{eq:gtail}
\end{equation}
The nonnegative coefficients of $\cA$ imply
$\|\cA\|_R=\cA(R)$. Together with $\|F_r\|_R\le B_R$,
this yields
\begin{equation}
    |F_r(t)-\cA(t)|
    \le \Bigl(\frac{\rho}{R}\Bigr)^r
        \bigl[B_R+\cA(R)\bigr].
    \label{eq:tail}
\end{equation}
Thus $F_r$ converges to $\cA$ uniformly on every disk
$|t|\le\rho<1$, exponentially fast in $r$. The uniform
absolute-weight bound ensures that the proliferation of
configurations at larger area deficits does not alter the
limiting amplitude.

Complete cancellation within the near-square sector is excluded
by a product formula for $\cA$. Jacobi's triple product at zero
argument, Eqs.~(20.5.3) and~(20.5.9) of~\cite{DLMF}, gives
\begin{equation}
    \sum_{j\in\Z}t^{j^2}
    =\prod_{n\ge1}
      \bigl(1-t^{2n}\bigr)
      \bigl(1+t^{2n-1}\bigr)^2.
    \label{eq:jacobi}
\end{equation}
Dividing by $\prod_{n\ge1}(1-t^n)^4$ and using
$1-t^{2n}=(1-t^n)(1+t^n)$, we obtain
\begin{equation}
    \cA(t)
    =\prod_{n\ge1}
      \frac{(1+t^n)(1+t^{2n-1})^2}{(1-t^n)^3}.
    \label{eq:product}
\end{equation}
For $|t|<1$, every factor is nonzero and
$\sum_{n\ge1}|t|^n<\infty$. Each numerator and denominator
product therefore converges to a finite nonzero limit,
establishing $\cA(t)\ne0$ throughout the unit disk.

At a fixed fugacity $\beta$ with $|\beta|>1$, choose
$\rho=|\beta|^{-1}$ and $R=|\beta|^{-1/2}$
in~\eqref{eq:tail}. Equation~\eqref{eq:Fr} then gives
\begin{equation}
    L_{2r}(\beta)
    =\beta^{r^2}F_r(1/\beta)
    =\beta^{r^2}
      \bigl[\cA(1/\beta)+\varepsilon_r\bigr],
    \label{eq:final}
\end{equation}
with
$|\varepsilon_r|\le[B_R+\cA(R)]\,|\beta|^{-r/2}$.
The near-square fluctuations thus contribute a nonzero amplitude
to the square weight, and the remaining droplets and clusters
do not change its leading growth, even at negative or complex
fugacity.

The operator $z\partial_z$, which enters the magnetization,
weights each contribution by its area exponent.
Differentiating~\eqref{eq:Fr} gives
\begin{equation}
    z\,\partial_zL_{2r}(z)
    =z^{r^2}\bigl[r^2F_r(t)-tF'_r(t)\bigr],
    \qquad t=1/z,
    \label{eq:magderiv}
\end{equation}
where the two terms account for the square area and the area
deficit, respectively. By~\eqref{eq:tail} and Cauchy's estimates,
$F'_r\to\cA'$ uniformly on smaller disks. The deficit contribution
$tF'_r(t)$ therefore remains bounded at $t=1/\beta$, while
$F_r(1/\beta)$ approaches the nonzero amplitude $\cA(1/\beta)$.
Hence
\begin{equation}
    \beta L'_{2r}(\beta)
    =r^2\beta^{r^2}
      \bigl[\cA(1/\beta)+o(1)\bigr].
    \label{eq:magfinal}
\end{equation}
Only $|\beta|>1$ entered these estimates, so~\eqref{eq:final}
and~\eqref{eq:magfinal} hold for every complex $b$ with $|b|>1$.
This proves Proposition~\ref{prop:asym}.

\textit{Proof of Theorem~\ref{thm:square}.}
At the conjugate fugacity, the coefficient growth is exponential
in the droplet area, whereas the expansion order is proportional
to its perimeter. For sufficiently large $r$,
$|\varepsilon_r|\le|\cA(1/\beta)|/2$,
and~\eqref{eq:final} yields
\begin{equation}
    \log|L_{2r}(\beta)|
    \ge r^2\log|\beta|
       +\log\frac{|\cA(1/\beta)|}{2}.
    \label{eq:lower}
\end{equation}
Since $\log|\beta|>0$, this quadratic growth contradicts
Maillet's bound~\eqref{eq:maillet}, which permits only
$\log^+|L_{2r}(\beta)|=O(r\log r)$.
Thus $\Phi_\beta$ is differentially transcendental. An algebraic
differential equation for $\Phi(u,\alpha)$ would transfer
coefficientwise to a nonzero equation for $\Phi_\beta$
by~\eqref{eq:conjDA}. Consequently, $\Phi(u,\alpha)$ is
differentially transcendental for every algebraic integer
$\alpha\in(0,1)$.

The physical pressure satisfies
$p=2K+H+\Phi(e^{-4K},e^{-2H})$.
Both the change of variable $u=e^{-4K}$ and its inverse
$K=-\frac14\log u$ are differentially algebraic, as is the
background term $2K+H$. Closure under composition and subtraction
therefore implies that an algebraic differential equation for $p$
on a low-temperature interval would give one for $\Phi(u,\alpha)$.
The pressure is accordingly differentially transcendental at every
$H>0$ with algebraic-integer fugacity.

For magnetization, joint analyticity of $\Phi$ in $(u,z)$ in the
convergent low-temperature region gives
$M=\partial_Hp=1-2z\partial_z\Phi$.
Its Taylor coefficients $-2zL'_m(z)$ belong to $\Q[z]$, so
conjugation applies as before. Their asymptotic
behavior~\eqref{eq:magfinal} also violates Maillet's bound:
the additional factor $r^2$ contributes only $O(\log r)$ to the
logarithm of the coefficient magnitude. This establishes
differential transcendence of $M$ at the same physical fugacities.

Joint real analyticity of the pressure in $(K,H)$ for $K>0$ and
$H\ne0$~\cite{Ott2020} implies real analyticity of both $p$ and $M$
in $K$ at fixed nonzero $H$. Any polynomial differential relation
valid on an open interval of $K>0$ would therefore extend throughout
the connected region $K>0$, including the low-temperature
region~\cite{LuoTian2026}. The obstruction consequently holds on
every nonempty open interval of positive coupling. Together with
the earlier result for fugacities that are not algebraic
integers~\cite{LuoTian2026} and spin reversal, this proves
Theorem~\ref{thm:square} for every nonzero real $H$. The closed
forms discussed in Sec.~\ref{sec:intro} are excluded by their
differential algebraicity.

\textit{Other fugacities.}
The argument also reaches complex fugacities.

\begin{corollary}
\label{cor:complex}
The formal series $\Phi(u,z)$ is differentially transcendental at every complex $z$ other than zero and the roots of unity.
\end{corollary}
By Kronecker's theorem, for every nonzero algebraic integer \(\alpha\) that is not a root of unity, Kronecker’s theorem supplies an embedding \(\sigma\) with \(\beta=\sigma(\alpha)\) and \(|\beta|>1\).~\cite{Kronecker1857}. For algebraic fugacities that are not algebraic integers, and for transcendental fugacities, the argument of~\cite{LuoTian2026} does not depend on $|z|<1$ either. This proves Corollary~\ref{cor:complex}. Outside the disk $|z|<1$ this concerns the formal series, without a general identification with an equilibrium pressure. At real $z>1$, for instance, it is the formal expansion about the all-plus state that the field disfavors. It is shown in~\cite{AdamczewskiDreyfusHardouin2021} that a power series obeying a linear $q$-difference equation with rational coefficients is either rational or differentially transcendental when $q$ is not a root of unity. This covers the series $S(u)=\sum_{r\ge0}z^{r^2}u^{2r}$ of the square droplets alone, which obeys $S(u)=1+zu^2S(zu)$, but not $\Phi$.

\section{Real conjugates in every dimension}
\label{sec:higher}
We now extend the result to every dimension $d\ge2$ at algebraic fugacities with a real conjugate $q>1$, where positivity replaces the control of cancellations on the square lattice by~\eqref{eq:Fstable},~\eqref{eq:keybound}, and~\eqref{eq:product}.
\begin{theorem}
\label{thm:real}
Let $d\ge2$. On $\Z^d$, the formal low-temperature series of the pressure obeys no algebraic differential equation in $u$ at every real fugacity $q>1$. Consequently, at every real $H\ne0$ at which $e^{-2|H|}$ is algebraic and has a real conjugate larger than one, the pressure and the magnetization on $\Z^d$ obey no algebraic differential equation in $K$ on any nonempty open interval of $K>0$.
\end{theorem}
To prove Theorem~\ref{thm:real}, we note that on the hypercubic lattice $\Z^d$ the Hamiltonian~\eqref{eq:Hamiltonian} gives the low-temperature expansion~\cite{ButeraPernici2016}
\begin{equation}
	p=dK+H+\Phi_d(u,z),\qquad \Phi_d(u,z)=\sum_{m\ge d}L^{(d)}_m(z)\,u^m.
	\label{eq:seriesd}
\end{equation}
A single reversed spin has $|\partial S|=2d$, and every nonempty $S$ has $|\partial S|\ge2d$, since its extreme sites in the $2d$ coordinate directions each have a boundary bond pointing out of $S$. Hence $m(S)=|\partial S|/2\ge d$, and the sum in~\eqref{eq:seriesd} starts at $m=d$. The construction of Sec.~\ref{sec:review} carries over~\cite{LuoTian2026}. The coefficients are finite sums over Mayer clusters,
\begin{equation}
	L^{(d)}_m(z)=\sum_{\mathcal C}\frac{1}{k!}\,\varphi^T(\gamma_1,\ldots,\gamma_k)\,z^{\sum_i|\gamma_i|},
	\label{eq:Lmd}
\end{equation}
where $\mathcal C$ runs over the translation classes of Mayer clusters of droplets of $\Z^d$ with $\sum_im(\gamma_i)=m$, and they obey $mL^{(d)}_m\in\Z[z]$. The series~\eqref{eq:seriesd} converges near $u=0$ for $|z|<1$, with a sum that is analytic in $(u,z)$ jointly~\cite{LuoTian2026}, and the pressure is real analytic in $(K,H)$ jointly for $K>0$ and $H\ne0$ by Corollary~1.4 of~\cite{Ott2020}. By Theorem~6 of~\cite{EllisFriedgutKindlerYehudayoff2016}, every droplet obeys
\begin{equation}
	|S|\le\Bigl(\frac{m(S)}{d}\Bigr)^{d/(d-1)}\le m(S)^2,
	\label{eq:isoperimetric}
\end{equation}
and the cube of side $r$, with $|S|=r^d$ and $m(S)=dr^{d-1}$, attains the first bound. For $d\ge3$ we know neither the analog of~\eqref{eq:stable} for near-cube droplets nor the analog of the polygon bound~\eqref{eq:polygons}. Nevertheless, when the fugacity has a real conjugate $q>1$, we need neither, because all droplet weights $q^{|S|}$ are then positive.

Let $q>1$ be real. We define $D_m(q)$ and $E_m(q)$ as in~\eqref{eq:DE}, now with the droplets of $\Z^d$, including holes and handles. At $z=q>0$ every single droplet enters~\eqref{eq:Lmd} with the positive weight $q^{|S|}$, so
\begin{equation}
	L^{(d)}_m(q)=D_m(q)+\sum_{\mathcal C:\,k\ge2}\frac{1}{k!}\,\varphi^T(\gamma_1,\ldots,\gamma_k)\,q^{\sum_i|\gamma_i|}.
	\label{eq:Lsplit}
\end{equation}
The last sum has modulus at most $E_m(q)$, and the cube is one droplet with $m=dr^{d-1}$ and weight $q^{r^d}$. Hence
\begin{equation}
	D_m(q)\le\bigl|L^{(d)}_m(q)\bigr|+E_m(q),\qquad D_{dr^{d-1}}(q)\ge q^{r^d}.
	\label{eq:positive}
\end{equation}
On $\Z^d$ a droplet $\gamma_j$ that is incompatible with $\gamma_i$ contains a site of $\gamma_i$ or one of its $2d$ neighbors, so it has at most $(2d+1)|\gamma_i||\gamma_j|\le(2d+1)m^4$ positions relative to $\gamma_i$, by~\eqref{eq:isoperimetric}, in place of $5m^4$. The tree bound and the argument that led to~\eqref{eq:Ebound} give
\begin{equation}
	E_m(q)\le\sum_{k=2}^{\lfloor m/2\rfloor}\frac{k^{k-2}}{k!}\bigl[(2d+1)m^4\bigr]^{k-1}\sum_{\{m_i\}}\prod_{i=1}^kD_{m_i}(q),
	\label{eq:triangular}
\end{equation}
where the inner sum runs over the compositions $m=m_1+\cdots+m_k$ with all $m_i\ge2$. Every part obeys $m_i\le m-2(k-1)<m$, so~\eqref{eq:positive} and~\eqref{eq:triangular} bound $D_m(q)$ by $|L^{(d)}_m(q)|$ and by the lower orders, and we need no analog of~\eqref{eq:polygons}.

Suppose that $\Phi_d(u,q)$ obeys a nonzero algebraic differential equation. By~\eqref{eq:mailletthm} applied to $1+\Phi_d(u,q)$, as in Sec.~\ref{sec:conjugate},
\begin{equation}
	\bigl|L^{(d)}_m(q)\bigr|\le C_0\,B^m\,(m!)^{s_0}
	\label{eq:mailletd}
\end{equation}
with constants $C_0$, $B$, and $s_0$. We fix an integer $s_1\ge\max\{s_0,3\}$ and show by induction
on $m$ that
\begin{equation}
    D_m(q)\le C_1^m(m!)^{s_1}
    \label{eq:indclaim}
\end{equation}
for a constant $C_1$. We first bound the factorials that enter the induction step. For $m_i\ge2$ with $\sum_im_i=m$, the log-convexity of $n!$ puts the largest product of factorials at $k-1$ parts equal to two, so
\begin{equation}
	\prod_{i=1}^km_i!\le2^{k-1}\,(m-2k+2)!.
	\label{eq:logconvex}
\end{equation}
Since $\log x$ increases,
\begin{equation}
	\sum_{j=m-n+1}^{m}\log j\ge\int_{m-n}^{m}\log x\,dx\ge n\log\frac{m}{e},
\end{equation}
so that $m!/(m-n)!\ge(m/e)^n$. With $n=2k-2$, this and~\eqref{eq:logconvex} give
\begin{equation}
	\frac{\prod_im_i!}{m!}\le\Bigl(\frac{2e^2}{m^2}\Bigr)^{k-1}.
	\label{eq:factorials}
\end{equation}
There are at most $m^{k-1}$ compositions of $m$ into $k$ parts, and $k^{k-2}/k!\le e^k$. If~\eqref{eq:indclaim} holds for all orders below $m$, then~\eqref{eq:triangular} gives
\begin{equation}
	\frac{E_m(q)}{C_1^m(m!)^{s_1}}\le\sum_{k\ge2}e^k\bigl[(2d+1)m^5\bigr]^{k-1}\max_{\{m_i\}}\Bigl(\frac{\prod_im_i!}{m!}\Bigr)^{s_1},
\end{equation}
and~\eqref{eq:factorials} turns this into
\begin{equation}
	\frac{E_m(q)}{C_1^m(m!)^{s_1}}\le e\sum_{k\ge2}\Bigl[e(2d+1)(2e^2)^{s_1}\,m^{5-2s_1}\Bigr]^{k-1}.
	\label{eq:induction}
\end{equation}
The right-hand side does not depend on $C_1$ and tends to zero as $m\to\infty$ because $s_1\ge3$. We choose $m_0$ such that it is at most $1/2$ for $m\ge m_0$, and then $C_1$ such that~\eqref{eq:indclaim} holds for $m<m_0$ and $C_0B^m(m!)^{s_0}\le C_1^m(m!)^{s_1}/2$ for all $m$. For $m\ge m_0$,~\eqref{eq:positive},~\eqref{eq:mailletd}, and~\eqref{eq:induction} then give
\begin{equation}
	D_m(q)\le\tfrac12C_1^m(m!)^{s_1}+\tfrac12C_1^m(m!)^{s_1}=C_1^m(m!)^{s_1},
\end{equation}
which completes the induction. At $m=dr^{d-1}$, the cube bound $D_m(q)\ge q^{r^d}$ in~\eqref{eq:positive} and~\eqref{eq:indclaim} give
\begin{equation}
	r^d\log q\le m\log C_1+s_1\log m!=O\bigl(r^{d-1}\log r\bigr),
	\label{eq:cubecontra}
\end{equation}
which is false for large $r$, because the left side grows like $r^d$ and the right side only like $r^{d-1}\log r$. This contradicts~\eqref{eq:mailletd}, the only consequence of the assumed equation that we used. Hence $\Phi_d(u,q)$ obeys no algebraic differential equation, for every $d\ge2$ and every real $q>1$.

Let $\alpha\in(0,1)$ be algebraic with a real conjugate $q=\sigma(\alpha)>1$. Since $L^{(d)}_m\in\Q[z]$, we have $\sigma(L^{(d)}_m(\alpha))=L^{(d)}_m(q)$, and the descent and conjugation of Sec.~\ref{sec:conjugate} map an algebraic differential equation for $\Phi_d(u,\alpha)$ to one for $\Phi_d(u,q)$, which we have excluded. By the analyticity of the pressure in $(K,H)$, the pressure on $\Z^d$ therefore obeys no algebraic differential equation on any interval of the coupling at the corresponding fixed field. The magnetization $M=\partial_Hp=1-2z\partial_z\Phi_d$ has the Taylor coefficients $-2z\,\partial_zL^{(d)}_m(z)\in\Q[z]$, and at $z=q$
\begin{equation}
\begin{split}
	q\,\partial_qL^{(d)}_m(q)&=D^{(1)}_m(q)\\
	&\quad+\sum_{\mathcal C:\,k\ge2}\frac{1}{k!}\,\varphi^T(\gamma_1,\ldots,\gamma_k)\Bigl(\sum_i|\gamma_i|\Bigr)q^{\sum_i|\gamma_i|},
\end{split}
	\label{eq:magd}
\end{equation}
with the positive single-droplet part
\begin{equation}
	D^{(1)}_m(q)=\sum_{S:\,m(S)=m}|S|\,q^{|S|}.
	\label{eq:D1}
\end{equation}
Since $\sum_i|\gamma_i|\le k\prod_i|\gamma_i|$, the part with several droplets obeys~\eqref{eq:triangular} with $D^{(1)}_{m_i}$ in place of $D_{m_i}$ and an extra factor $k\le2^k$. The induction goes through unchanged for $D^{(1)}_m$, and the cube gives $D^{(1)}_{dr^{d-1}}(q)\ge r^dq^{r^d}$. Hence the magnetization obeys no algebraic differential equation either. This proves Theorem~\ref{thm:real}.

The algebraic fugacities $\alpha\in(0,1)$ with a real conjugate $\beta = q=\sigma(\alpha)>1$ in~\eqref{eq:conjugate}, are dense in $(0,1)$. To see this, note that for $N\ge3$ and $1\le M\le N-2$, the polynomial $P_{N,M}(x)=x^2-Nx+M$ obeys
\begin{equation}
	P_{N,M}(0)=M>0,\qquad P_{N,M}(1)=1-N+M<0,
\end{equation}
so its smaller root $\alpha_{N,M}=(N-\sqrt{N^2-4M})/2$ lies in $(0,1)$, and its other root $N-\alpha_{N,M}$ is larger than one. A rational root of the monic polynomial $P_{N,M}\in\Z[x]$ would be an integer, so $P_{N,M}$ is irreducible over $\Q$, and $N-\alpha_{N,M}$ is a conjugate of $\alpha_{N,M}$. Given any $t\in(0,1)$, we take $M=\lfloor Nt\rfloor$, which obeys $1\le M\le N-2$ for large $N$. Dividing $P_{N,M}(\alpha_{N,M})=0$ by $N$ gives $\alpha_{N,M}=M/N+\alpha_{N,M}^2/N$, and hence
\begin{equation}
	|\alpha_{N,M}-t|\le\frac{\alpha_{N,M}^2}{N}+\Bigl|\frac{\lfloor Nt\rfloor}{N}-t\Bigr|\le\frac2N,
	\label{eq:dense}
\end{equation}
so $\alpha_{N,M}\to t$ as $N\to\infty$, and every point of $(0,1)$ is a limit of such fugacities.

\textit{Coupled layers.} We also use Theorem~\ref{thm:square} to treat higher-dimensional models with a generic interlayer coupling. On $\Z^d$ with $d\ge3$, we keep the coupling $J$ on the bonds along $e_1$ and $e_2$ and put $\lambda J$ with $\lambda>0$ on the bonds along the other axes,
\begin{equation}
\begin{split}
	\mathcal H_\lambda=&-J\sum_{\langle ij\rangle\parallel e_1,e_2}\sigma_i\sigma_j\\
	&-\lambda J\sum_{\langle ij\rangle\parallel e_3,\ldots,e_d}\sigma_i\sigma_j-h\sum_i\sigma_i.
\end{split}
	\label{eq:layers}
\end{equation}
Here $\langle ij\rangle\parallel e_a$ means $j-i=\pm e_a$, so each spin has four neighbors in its layer, joined by bonds of coupling $J$, and $2(d-2)$ neighbors in the adjacent layers, joined by bonds of coupling $\lambda J$.
\begin{corollary}
\label{cor:layers}
Let $d\ge3$, let $H\ne0$ be real, and let $\lambda>0$ be transcendental over $\Q(e^{-2H})$. Then the pressure of~\eqref{eq:layers} obeys no algebraic differential equation in $K$ on any nonempty open interval of $K>0$.
\end{corollary}
To see this, we write $\lambda_{ij}\in\{1,\lambda\}$ for the relative coupling of the bond $\langle ij\rangle$, $\mathcal E_N=\sum_{\langle ij\rangle}\lambda_{ij}\sigma_i\sigma_j$ for the bond energy of a box with $N$ sites, and $\langle\cdot\rangle_H$ for the product measure of independent spins in the field $H$, so that
\begin{equation}
	\langle\sigma_i\rangle_H=\tanh H=\frac{1-z}{1+z}.
	\label{eq:indep}
\end{equation}
At fixed $H$, the high-temperature expansion
\begin{equation}
\begin{split}
	p-\log(2\cosh H)&=\lim_{N\to\infty}\frac1N\log\bigl\langle e^{K\mathcal E_N}\bigr\rangle_H\\
	&=\sum_{n\ge1}a_n(\lambda,z)\,K^n
\end{split}
	\label{eq:hightemp}
\end{equation}
converges near $K=0$, uniformly in $N$. To see this, we write $e^{K\lambda_{ij}\sigma_i\sigma_j}=\cosh(K\lambda_{ij})\bigl[1+\sigma_i\sigma_j\tanh(K\lambda_{ij})\bigr]$ and expand the product over the bonds. Since $\sigma_i^2=1$ and the spins are independent, a set $\gamma$ of bonds gives $\bigl\langle\prod_{\langle ij\rangle\in\gamma}\sigma_i\sigma_j\bigr\rangle_H=(\tanh H)^{o(\gamma)}$ by~\eqref{eq:indep}, where $o(\gamma)$ is the number of sites that belong to an odd number of bonds of $\gamma$. Hence $\langle e^{K\mathcal E_N}\rangle_H\big/\prod_{\langle ij\rangle}\cosh(K\lambda_{ij})$ is the partition function of a gas of connected sets $\gamma$ of bonds that share no site, with the activities
\begin{equation}
	w(\gamma)=(\tanh H)^{o(\gamma)}\prod_{\langle ij\rangle\in\gamma}\tanh(K\lambda_{ij}),
	\label{eq:htpolymer}
\end{equation}
and its cluster expansion converges for small $|K|$, uniformly in $N$~\cite{FernandezProcacci2007}. The cumulants $\kappa_n(\mathcal E_N)$ of the bond energy under $\langle\cdot\rangle_H$ are defined by
\begin{equation}
	\log\bigl\langle e^{K\mathcal E_N}\bigr\rangle_H=\sum_{n\ge1}\frac{K^n}{n!}\,\kappa_n(\mathcal E_N),
	\label{eq:cumulants}
\end{equation}
so that $a_n=\lim_{N\to\infty}\kappa_n(\mathcal E_N)/(n!\,N)$. With $X_b=\sigma_i\sigma_j$ and $\lambda_b=\lambda_{ij}$ for the bond $b=\langle ij\rangle$, so that $\mathcal E_N=\sum_b\lambda_bX_b$, the cumulant is multilinear,
\begin{equation}
	\kappa_n(\mathcal E_N)=\sum_{b_1,\ldots,b_n}\lambda_{b_1}\cdots\lambda_{b_n}\,\kappa\bigl(X_{b_1},\ldots,X_{b_n}\bigr),
	\label{eq:multilinear}
\end{equation}
with the joint cumulant $\kappa(X_{b_1},\ldots,X_{b_n})=\partial_{t_1}\cdots\partial_{t_n}\log\langle e^{t_1X_{b_1}+\cdots+t_nX_{b_n}}\rangle_H$ at $t=0$. Each joint cumulant is a polynomial in $\tanh H$, since $\sigma_i^2=1$ and $\langle\sigma_i\rangle_H=\tanh H$. We call $b_1,\ldots,b_n$ connected if they cannot be split into two nonempty groups that share no site. A joint cumulant vanishes unless $b_1,\ldots,b_n$ are connected, because two such groups involve disjoint, hence independent, sets of spins. Hence $a_n(\lambda,z)$ is a finite sum over the translation classes of connected sequences $b_1,\ldots,b_n$, a polynomial of degree at most $n$ in $\lambda$ with coefficients in $\Q(z)$ by~\eqref{eq:indep}. At $\lambda=0$ only the sequences inside one layer remain, and
\begin{equation}
	\sum_{n\ge1}a_n(0,z)\,K^n=p_\square(K,H)-\log(2\cosh H)
\end{equation}
is the expansion of the square-lattice pressure $p_\square$. It obeys no algebraic differential equation, because it converges to $p_\square-\log(2\cosh H)$ on an interval $0<K<K_0$, on which $p_\square$ obeys none by Theorem~\ref{thm:square}. As in Sec.~\ref{sec:review}, we fix the finitely many monomials of a hypothetical equation
\begin{equation}
	\sum_{\mu,\nu}c_{\mu,\nu}\,K^{\mu}\prod_{j=0}^{s}\bigl(\partial_K^jf_\lambda\bigr)^{\nu_j}=0
	\label{eq:DAK}
\end{equation}
for the series $f_\lambda(K)=\sum_na_n(\lambda,z)K^n$, with complex coefficients $c_{\mu,\nu}$ that are not all zero. The coefficient of each power of $K$ in~\eqref{eq:DAK} is linear in the vector $c=(c_{\mu,\nu})$, so the equation becomes the infinite linear system
\begin{equation}
	\sum_{\mu,\nu}\mathsf M_{n,(\mu,\nu)}(\lambda)\,c_{\mu,\nu}=0,\qquad n\ge0,
	\label{eq:minorsystem}
\end{equation}
with the entries
\begin{equation}
	\mathsf M_{n,(\mu,\nu)}(\lambda)=[K^n]\,K^\mu\prod_{j=0}^{s}\bigl(\partial_K^jf_\lambda\bigr)^{\nu_j}\in\Q(z)[\lambda].
\end{equation}
At $\lambda=0$ its only solution is $c=0$, so some rows $n_1,n_2,\ldots$, as many as there are monomials $(\mu,\nu)_1,(\mu,\nu)_2,\ldots$, give a square minor
\begin{equation}
	D(\lambda)=\det\bigl[\mathsf M_{n_a,(\mu,\nu)_b}(\lambda)\bigr]_{a,b}\in\Q(z)[\lambda],\qquad D(0)\ne0.
	\label{eq:minorD}
\end{equation}
This nonzero polynomial vanishes only at numbers $\lambda$ that are algebraic over $\Q(z)$. For every $\lambda$ that is transcendental over $\Q(z)$, the system~\eqref{eq:minorsystem} has only the solution $c=0$ for every choice of monomials, and the high-temperature expansion of the coupled layers therefore obeys no algebraic differential equation. By Corollary~1.4 and Remark~2 of~\cite{Ott2020}, which states that the proof extends to finite-range interactions, the pressure of the coupled layers is real analytic in $K>0$ at fixed $H\ne0$. This proves Corollary~\ref{cor:layers}, whose possible exceptions lie in the countable set of roots of nonzero polynomials in $\Q(e^{-2H})[x]$, which may depend on $H$. A fixed $\lambda$ that is transcendental over $\Q$ thus covers every algebraic fugacity, but not necessarily every transcendental one, since $\lambda$ is algebraic over $\Q(z)$ for some transcendental $z$. The $\lambda=1$ case is not covered.

\section{Decorated lattices}
\label{sec:decorated}

Decorated lattices carry extra spins, and summing over them gives an effective square lattice whose coupling and fugacity both depend on the original coupling. Such lattices are exactly solvable in a nonzero field when the effective field vanishes, as for the super-exchange antiferromagnet of~\cite{Fisher1960}, whose field acts only on the decorating spins, and for mixed-spin decorations of the Lieb lattice~\cite{StreckaKarlova2026}. With uniform coupling and field the effective field does not vanish, and at fixed field the effective fugacity varies with the temperature, so Theorem~\ref{thm:square} does not apply directly, but the near-square droplets still control the large-order coefficients.
\begin{theorem}
\label{thm:decorated}
Let $\kappa\ge2$, replace every bond of the square lattice by a chain of $\kappa$ bonds and $\kappa-1$ spins, and give all bonds the coupling $J>0$ and all spins the field $h$. Then, at every real $H\ne0$, the pressure obeys no algebraic differential equation in $K$ on any nonempty open interval of $K>0$.
\end{theorem}

\subsection{Exact reduction on the Lieb lattice}

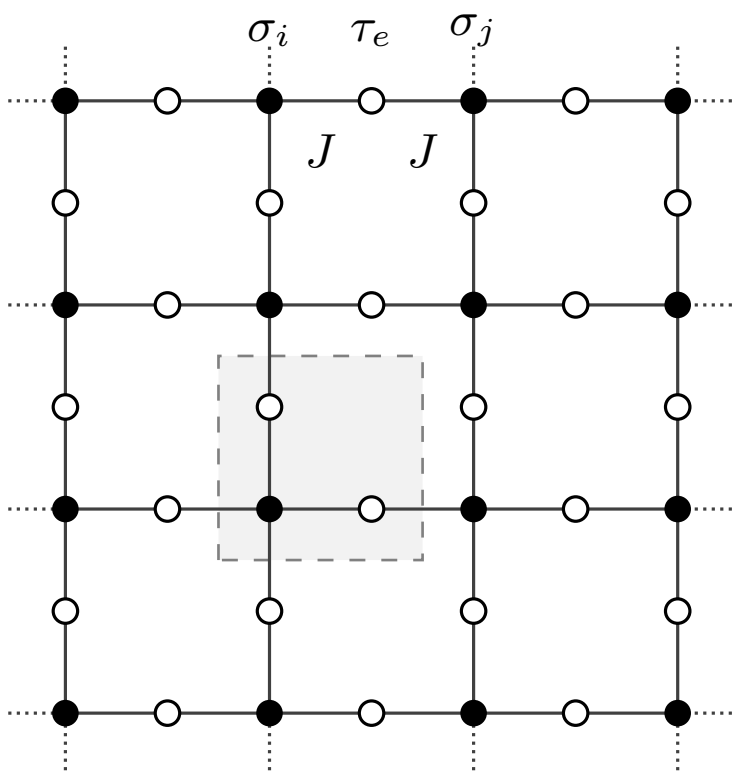
\begin{figure}[t]
\centering
\begin{tikzpicture}[
  x=1.35cm, y=1.35cm,
  bond/.style={draw=black!75, line width=0.6pt},
  sigma site/.style={
    circle, fill=black, draw=black,
    inner sep=0pt, minimum size=4.6pt
  },
  tau site/.style={
    circle, fill=white, draw=black,
    line width=0.6pt,
    inner sep=0pt, minimum size=4.6pt
  },
  every node/.style={font=\small}
]

% Primitive cell containing one sigma and two tau sites.
\draw[fill=black!5, draw=black!50, dashed, line width=0.5pt]
  (0.75,0.75) rectangle (1.75,1.75);

% Bonds and continuations of the infinite lattice.
\foreach \n in {0,...,3} {
  \draw[bond] (0,\n) -- (3,\n);
  \draw[bond] (\n,0) -- (\n,3);
  \draw[bond,densely dotted] (-0.28,\n) -- (0,\n);
  \draw[bond,densely dotted] (3,\n) -- (3.28,\n);
  \draw[bond,densely dotted] (\n,-0.28) -- (\n,0);
  \draw[bond,densely dotted] (\n,3) -- (\n,3.28);
}

% Original square-lattice sites.
\foreach \x in {0,...,3}
  \foreach \y in {0,...,3}
    \node[sigma site] at (\x,\y) {};

% Decorating sites on horizontal bonds.
\foreach \x in {0,...,2}
  \foreach \y in {0,...,3}
    \node[tau site] at ({\x+0.5},\y) {};

% Decorating sites on vertical bonds.
\foreach \x in {0,...,3}
  \foreach \y in {0,...,2}
    \node[tau site] at (\x,{\y+0.5}) {};

% Label one decorated bond.
\node[above=7pt] at (1,3)   {$\sigma_i$};
\node[above=7pt] at (1.5,3) {$\tau_e$};
\node[above=7pt] at (2,3)   {$\sigma_j$};
\node[below=3pt] at (1.25,3) {$J$};
\node[below=3pt] at (1.75,3) {$J$};

\end{tikzpicture}
\caption{Lieb lattice, obtained by inserting a decorating spin $\tau_e$ (open circle) on each bond $e=\langle ij\rangle$ of the square lattice, whose original spins $\sigma_i$ are shown as filled circles. Each original bond is replaced by the two bonds $\sigma_i$--$\tau_e$ and $\tau_e$--$\sigma_j$, both with coupling $J>0$; the field $h$ acts uniformly on all spins. The dashed primitive cell contains one $\sigma$ spin and two $\tau$ spins, with four interaction bonds per cell. This is the $\kappa=2$ chain decoration.}
\label{fig:lieb-lattice}
\end{figure}

The Lieb lattice~\cite{Lieb1989} (shown in Fig.~\ref{fig:lieb-lattice}) corresponds to $\kappa=2$: a spin $\tau_e$ is inserted on every square-lattice bond $e=\langle ij\rangle$. Its Hamiltonian is
\begin{equation}
    \mathcal H_L =-J\sum_{e=\langle ij\rangle}\tau_e(\sigma_i+\sigma_j) -h\Bigl(\sum_i\sigma_i+\sum_e\tau_e\Bigr).
\label{eq:Lieb}
\end{equation}
Summing over each $\tau_e$ gives an interaction between the retained spins $\sigma_i$ and $\sigma_j$, together with a contribution to their local fields. This is the decoration-iteration transformation~\cite{Fisher1959}. With the aligned-state normalization, the bond weight is
\begin{equation}
    w_{\sigma_i\sigma_j}
    =e^{-2K-H}\sum_{\tau=\pm1}
      e^{\tau[K(\sigma_i+\sigma_j)+H]},
    \label{eq:wdef}
\end{equation}
so that
\begin{equation}
    w_{++}=1+uz,\qquad
    w_{--}=z+u,\qquad
    w_{+-}=\sqrt u\,(1+z),
    \label{eq:decoration}
\end{equation}
with $w_{-+}=w_{+-}$.

The effective coupling and field follow by separating the boundary and area weights of a reversed domain. On a torus with $N$ retained spins, let $S$ contain $n$ reversed spins and have half-perimeter $m$. If $b_{++}$, $b_{--}$, and $b_{+-}$ count the three types of square-lattice bonds, then
\begin{equation}
    b_{+-}=2m,\qquad
    4n=2b_{--}+b_{+-},\qquad
    b_{++}+b_{+-}+b_{--}=2N.
    \label{eq:bondcount}
\end{equation}
Thus $b_{--}=2n-m$ and $b_{++}=2N-2n-m$, and the normalized partition function is
\begin{equation}
    P_{L,N}(u,z)
    =\sum_S z^n
       w_{++}^{2N-2n-m}w_{--}^{2n-m}w_{+-}^{2m}.
    \label{eq:LiebP}
\end{equation}
The weight of $S$ factors as
\begin{equation}
\begin{split}
    &z^nw_{++}^{2N-2n-m}w_{--}^{2n-m}w_{+-}^{2m}\\
    &\qquad=w_{++}^{2N}
       \Bigl(\frac{z\,w_{--}^2}{w_{++}^2}\Bigr)^n
       \Bigl(\frac{w_{+-}^2}{w_{++}w_{--}}\Bigr)^m.
\end{split}
    \label{eq:collect}
\end{equation}
The factor raised to $n$ is the effective fugacity, while the factor raised to $m$ is the effective boundary weight. Consequently,
\begin{gather}
    P_{L,N}(u,z)=(1+uz)^{2N}P_{\square,N}(U,Z),
    \nonumber\\
    U=\frac{u(1+z)^2}{(z+u)(1+uz)},\qquad
    Z=\frac{z(z+u)^2}{(1+uz)^2},
    \label{eq:Liebmap}
\end{gather}
where $P_{\square,N}(U,Z)=\sum_SU^{m(S)}Z^{|S|}$.

Each cell contains three spins and four bonds. In the physical low-temperature region, restoring the aligned-state weight $e^{N(4K+3H)}$ and taking the thermodynamic limit gives the pressure per spin,
\begin{equation}
    p_L=\lim_{N\to\infty}\frac{\log Z_{L,N}}{3N}
       =\frac43K+H+\Phi_L(u,z),
    \label{eq:pL}
\end{equation}
with
\begin{equation}
    \Phi_L(u,z)
    =\frac23\log(1+uz)+\frac13\Phi(U,Z).
    \label{eq:LiebPhi}
\end{equation}
At nonphysical fugacities, we use~\eqref{eq:LiebPhi} as an identity of formal low-temperature series. At zero field,
\begin{equation}
    U=\frac{4u}{(1+u)^2},\qquad Z=1,
    \label{eq:Liebzero}
\end{equation}
and Onsager's solution applies~\cite{Onsager1944}. At nonzero field, the dependence of $Z$ on $u$ mixes different orders of the square-lattice expansion. The required obstruction must therefore be established for the composed series $\Phi(U(u),Z(u))$.

\subsection{Stability of the droplet asymptotics}

The leading growth survives analytic changes of the coupling and fugacity because mixing in lower orders reduces the maximal droplet area. Let $U(u)=cu+O(u^2)$ and $Z(u)=b+O(u)$ be analytic near $u=0$, with $c\ne0$ and $|b|>1$. Define the formal series
\begin{equation}
    G(u)=\sum_{m\ge2}L_m\bigl(Z(u)\bigr)U(u)^m.
    \label{eq:Gdef}
\end{equation}
Every coefficient is a finite sum, since $U$ has order one. We prove
\begin{equation}
    [u^{2r}]G
    =c^{2r}b^{r^2}\bigl[\cA(1/b)+o(1)\bigr].
    \label{eq:substitution}
\end{equation}
The factor $c^{2r}$ changes the boundary weight, while $b^{r^2}$ retains the area growth of a square droplet.

Put $q=|b|$ and $W=U/(cu)$. Since $Z(0)=b$ and $W(0)=1$, there is a constant $C$ such that
\begin{equation}
    |Z(u)|\le q\,e^{C|u|},\qquad
    |W(u)|\le e^{C|u|}
    \label{eq:ZWbound}
\end{equation}
near the origin. The absolute-weight bound~\eqref{eq:keybound} gives
\begin{equation}
    \sum_j\bigl|[z^j]L_m\bigr|q^j
    \le D_m(q)+E_m(q)
    \le C''_q q^{A_m}.
    \label{eq:Lmq}
\end{equation}
Since $\deg L_m\le A_m$, it follows that
\begin{equation}
    \bigl|L_m\bigl(Z(u)\bigr)W(u)^m\bigr|
    \le C''_q q^{A_m}e^{C(A_m+m)|u|}.
\end{equation}
Choose $\eta>0$ small enough that all circles $|u|=\eta/m^2$, $m\ge2$, lie in this neighborhood. The scale $m^{-2}$ compensates for the maximal area $A_m=O(m^2)$. Cauchy's estimate then yields
\begin{equation}
    \bigl|[u^k]\{L_m(Z)W^m\}\bigr|
    \le C' q^{A_m}\Bigl(\frac{m^2}{\eta}\Bigr)^k,
    \qquad C'=C''_qe^{C\eta}.
    \label{eq:cauchyk}
\end{equation}

Using $U^m=c^mu^mW^m$, we have
\begin{equation}
    [u^n]G
    =\sum_{m=2}^{n}c^m
       [u^{n-m}]\{L_m(Z)W^m\}.
    \label{eq:Gexpand}
\end{equation}
At $n=2r$, the term $m=n$ is $c^nL_n(b)$. A term with $m=n-k$ acquires $k$ additional powers of $u$ from the substitution, but its maximal area is smaller by $A_n-A_{n-k}\ge nk/4$. For sufficiently large even $n$, equation~\eqref{eq:cauchyk} gives
\begin{equation}
\begin{split}
    &\sum_{k=1}^{n-2}
      \bigl|c^{n-k}[u^k]\{L_{n-k}(Z)W^{n-k}\}\bigr|\\
    &\qquad\le C'|c|^nq^{A_n}
      \sum_{k\ge1}\theta_n^k,
    \qquad
    \theta_n=\frac{n^2q^{-n/4}}{\eta|c|}.
\end{split}
    \label{eq:subtail}
\end{equation}
For every fixed $q>1$, $\theta_n\to0$. The loss of area thus suppresses all lower-order contributions, despite the mixing generated by the analytic substitution. Applying~\eqref{eq:final} to the remaining term $c^nL_n(b)$ proves~\eqref{eq:substitution}.

\subsection{Differential transcendence on the Lieb lattice}

The low-temperature limit of the decoration map is
\begin{equation}
    U=c(z)u+O(u^2),\qquad
    Z=z^3+O(u),\qquad c(z)=\frac{(1+z)^2}{z}.
    \label{eq:Liebcz}
\end{equation}
The power $z^3$ reflects the three spins per cell in the droplet interior; the boundary contribution is contained in $c(z)$.

Let $\alpha\in(0,1)$ be an algebraic integer and choose a conjugate $\beta$ with $|\beta|>1$. Since $c(\beta)\ne0$, equation~\eqref{eq:substitution} applies with $b=\beta^3$. The local term in~\eqref{eq:LiebPhi} has coefficients
\begin{equation}
    [u^n]\,\frac23\log(1+u\beta)
    =\frac{2(-1)^{n+1}\beta^n}{3n},
\end{equation}
whose growth is at most exponential in $n$. Its exponential growth in $r$ cannot cancel the droplet contribution, which grows exponentially in $r^2$. Hence
\begin{equation}
    [u^{2r}]\Phi_L(u,\beta)
    =\frac13c(\beta)^{2r}\beta^{3r^2}
      \bigl[\cA(\beta^{-3})+o(1)\bigr].
    \label{eq:Liebasym}
\end{equation}
Write $\Phi_L(u,z)=\sum_{n\ge1}\hat L_n(z)u^n$. The coefficients arise from the droplet expansion on the Lieb lattice and belong to $\Q[z]$. Descent and conjugation therefore apply as in Sec.~\ref{sec:conjugate}. Since $\cA(\beta^{-3})\ne0$,
\begin{equation}
    \log|\hat L_{2r}(\beta)|
    =3r^2\log|\beta|+2r\log|c(\beta)|+O(1),
\end{equation}
in contradiction with Maillet's bound~\eqref{eq:maillet}. The formal series at $\alpha$ is consequently differentially transcendental.

At algebraic fugacities that are not algebraic integers, the corresponding obstruction is denominator growth. Choose a valuation $v$ with $v(\alpha)<0$, normalized to extend $v_\ell$, and set $|x|_v=\ell^{-v(x)}$ and $q=|\alpha^3|_v>1$. At $z=\alpha$, the normalized substitution factors are
\begin{equation}
    W=\frac{1}{(1+u/\alpha)(1+\alpha u)},\qquad
    \frac{Z}{\alpha^3}
    =\frac{(1+u/\alpha)^2}{(1+\alpha u)^2}.
    \label{eq:WZalpha}
\end{equation}
Their coefficients of $u^j$ have absolute value at most $C^j$, with $C=|\alpha|_v$. Since $mL_m\in\Z[z]$, the ultrametric inequality gives
\begin{equation}
    \bigl|[u^k]\{L_m(Z)W^m\}\bigr|_v
    \le m\,q^{A_m}C^k.
    \label{eq:vbound}
\end{equation}
For $n=2r$ sufficiently large, Sec.~\ref{sec:review} gives $|L_n(\alpha^3)|_v=q^{r^2}$. Writing $c=c(\alpha)$, the relative size of every lower-order term in~\eqref{eq:Gexpand} is bounded by
\begin{equation}
    \frac{\bigl|c^m[u^{n-m}]\{L_m(Z)W^m\}\bigr|_v}
         {|c|_v^nq^{r^2}}
    \le n\Bigl(\frac{Cq^{-n/4}}{|c|_v}\Bigr)^{n-m}.
    \label{eq:vsuppress}
\end{equation}
This is strictly less than one for all $m<n$ when $r$ is large. The term $m=n$ therefore fixes the valuation of the sum:
\begin{equation}
    v([u^{2r}]G)=2r\,v(c)+3r^2v(\alpha).
\end{equation}
Here $v(c)=v(\alpha)$, because $v(1+\alpha)=v(\alpha)$. The local logarithm has
\begin{equation}
    v\Bigl([u^n]\frac23\log(1+u\alpha)\Bigr)
    =n\,v(\alpha)+v(2)-v(3n),
\end{equation}
which is larger than the valuation of the droplet term at $n=2r$ for large $r$. Including the factor $1/3$ in~\eqref{eq:LiebPhi} gives
\begin{equation}
    v([u^{2r}]\Phi_L(u,\alpha))
    =(3r^2+2r)v(\alpha)-v(3).
    \label{eq:Liebval}
\end{equation}
This violates the linear lower bound on valuations of differentially algebraic coefficients~\cite{SibuyaSperber1981}. Since $\hat L_n\in\Q[z]$, the specialization argument of Sec.~\ref{sec:review} also excludes transcendental fugacities.

The passage from the formal series to the pressure on every coupling interval requires analyticity along the physical decoration curve. For $0<u,z<1$,
\begin{equation}
\begin{split}
    (z+u)(1+uz)-u(1+z)^2&=z(1-u)^2>0,\\
    (1+uz)-(z+u)&=(1-u)(1-z)>0.
\end{split}
    \label{eq:Liebpositive}
\end{equation}
Hence
\begin{equation}
\begin{split}
    U&=1-\frac{z(1-u)^2}{(z+u)(1+uz)}\in(0,1),\\
    Z&=z\Bigl(\frac{z+u}{1+uz}\Bigr)^2\in(0,z).
\end{split}
    \label{eq:LiebUZ}
\end{equation}
The effective model remains ferromagnetic, and the eliminated spins reinforce the field acting on the retained spins:
\begin{equation}
    K_{\rm eff}=-\frac14\log U>0,\qquad
    H_{\rm eff}=-\frac12\log Z>H>0.
    \label{eq:Keff}
\end{equation}
Both effective parameters are real analytic in $K>0$ at fixed $H>0$. Joint analyticity of the square-lattice pressure~\cite{Ott2020} therefore implies analyticity of
\begin{equation}
\begin{split}
    p_L&=\frac43K+H+\frac23\log(1+uz)\\
       &\quad+\frac13\bigl[
       p_\square(K_{\rm eff},H_{\rm eff})
       -2K_{\rm eff}-H_{\rm eff}\bigr].
\end{split}
    \label{eq:pLanalytic}
\end{equation}
An algebraic differential relation on any open interval of $K>0$ would extend to the low-temperature region, contradicting the coefficient bounds above. Spin reversal gives the same conclusion for $H<0$.

\subsection{Chains of arbitrary length}

For a chain of $\kappa$ bonds, let $\tau_0=s$ and $\tau_\kappa=t$ be the retained endpoint spins. Summing over the $\kappa-1$ internal spins gives the normalized weight
\begin{equation}
\begin{split}
    a_{st}
    &=e^{-\kappa K-(\kappa-1)H}\\
    &\quad\times
      \sum_{\tau_1,\ldots,\tau_{\kappa-1}}
      \exp\Bigl(
      K\sum_{i=1}^{\kappa}\tau_{i-1}\tau_i
      +H\sum_{i=1}^{\kappa-1}\tau_i\Bigr).
\end{split}
    \label{eq:chainweight}
\end{equation}
Like endpoints admit an aligned chain with no broken bond, whereas unlike endpoints require an interface. Each broken bond contributes $\sqrt u$, and each reversed internal spin contributes $z$. Thus $a_{++}$ and $a_{--}$ are polynomials in $u,z$ with
\begin{equation}
    a_{++}=1+O(u),\qquad
    a_{--}=z^{\kappa-1}+O(u).
    \label{eq:chainlike}
\end{equation}
For unlike endpoints, the leading term sums over the $\kappa$ possible positions of a single interface:
\begin{equation}
    a_{+-}=\sqrt u\,S_\kappa(z)+O(u^{3/2}),\qquad
    S_\kappa(z)=\sum_{i=0}^{\kappa-1}z^i.
    \label{eq:Sell}
\end{equation}
Reflection of the chain gives $a_{-+}=a_{+-}$.

The bond counting~\eqref{eq:bondcount} applies unchanged, yielding $P_{\kappa,N}=a_{++}^{2N}P_{\square,N}(U_\kappa,Z_\kappa)$, where
\begin{equation}
    U_\kappa=\frac{a_{+-}^2}{a_{++}a_{--}},\qquad
    Z_\kappa=z\,\frac{a_{--}^2}{a_{++}^2}.
    \label{eq:chainmap}
\end{equation}
There are $n_\kappa=2\kappa-1$ spins and $2\kappa$ bonds per cell. Accordingly, $p_\kappa=(2\kappa/n_\kappa)K+H+\Phi_\kappa(u,z)$, with
\begin{equation}
    \Phi_\kappa(u,z)
    =\frac{2}{n_\kappa}\log a_{++}(u,z)
     +\frac{1}{n_\kappa}\Phi(U_\kappa,Z_\kappa).
    \label{eq:chainPhi}
\end{equation}
At $\kappa=2$, these formulas reduce to the Lieb-lattice map. At zero field, spin reversal gives $a_{--}=a_{++}$ and hence $Z_\kappa=1$, so the pressure again follows from Onsager's solution. For nonzero fixed $z$,~\eqref{eq:chainlike} and~\eqref{eq:Sell} together give
\begin{equation}
    U_\kappa=c_\kappa(z)u+O(u^2),\qquad
    Z_\kappa=z^{n_\kappa}+O(u),
    \label{eq:chainUZ}
\end{equation}
where $c_\kappa(z)=S_\kappa(z)^2z^{1-\kappa}$. The cell spin count $n_\kappa$ multiplies the droplet-area exponent, while $c_\kappa$ changes its boundary weight.

Since
\begin{equation}
    S_\kappa(z)=\frac{z^\kappa-1}{z-1}
      =\prod_{j=1}^{\kappa-1}
       \bigl(z-e^{2\pi ij/\kappa}\bigr),
    \label{eq:Sroots}
\end{equation}
$S_\kappa(\beta)\ne0$ for $|\beta|>1$, and hence $c_\kappa(\beta)\ne0$. The local logarithm in~\eqref{eq:chainPhi} has at most exponentially growing coefficients, because $a_{++}(u,\beta)$ is a polynomial with constant term one. The substitution estimate~\eqref{eq:substitution} therefore gives
\begin{equation}
    [u^{2r}]\Phi_\kappa(u,\beta)
    =\frac{1}{n_\kappa}
      c_\kappa(\beta)^{2r}\beta^{n_\kappa r^2}
      \bigl[\cA(\beta^{-n_\kappa})+o(1)\bigr],
    \label{eq:chainasym}
\end{equation}
and
\begin{equation}
\begin{split}
    \log\bigl|[u^{2r}]\Phi_\kappa(u,\beta)\bigr|
    &=n_\kappa r^2\log|\beta|\\
    &\quad+2r\log|c_\kappa(\beta)|+O(1).
\end{split}
\end{equation}
The coefficients of $\Phi_\kappa$ belong to $\Q[z]$ by the droplet expansion on the decorated lattice. Conjugation and Maillet's bound exclude every algebraic-integer fugacity in $(0,1)$.

For a valuation with $v(\alpha)<0$, the highest power of $\alpha$ in $S_\kappa(\alpha)$ has strictly smallest valuation, so
\begin{equation}
\begin{split}
    v(S_\kappa(\alpha))&=(\kappa-1)v(\alpha),\\
    v(c_\kappa(\alpha))&=(\kappa-1)v(\alpha).
\end{split}
\end{equation}
The normalized factors $a_{++}(u,\alpha)$, $a_{--}(u,\alpha)/\alpha^{\kappa-1}$, and $a_{+-}(u,\alpha)/[\sqrt u\,S_\kappa(\alpha)]$ are polynomials with constant term one. Their coefficients have $v$-adic absolute values bounded by $C^j$ at order $u^j$ for some fixed $C\ge1$. For series with constant term one, this bound is preserved under multiplication, inversion, and nonnegative integer powers. It therefore holds for $W_\kappa=U_\kappa/[c_\kappa(\alpha)u]$ and $Z_\kappa/\alpha^{n_\kappa}$. The estimates~\eqref{eq:vbound}--\eqref{eq:vsuppress} apply with $q=|\alpha^{n_\kappa}|_v>1$. The term inherited from $L_{2r}(\alpha^{n_\kappa})$ again determines the valuation. The local logarithm satisfies $|[u^n]\log a_{++}|_v\le nC^n$, so its valuation is bounded below linearly in $n$ and cannot cancel the quadratic droplet contribution. Thus, for sufficiently large $r$,
\begin{equation}
\begin{split}
    v([u^{2r}]\Phi_\kappa(u,\alpha))
    &=2r\,v(c_\kappa)
      +n_\kappa r^2v(\alpha)-v(n_\kappa)\\
    &=\bigl[n_\kappa r^2+2r(\kappa-1)\bigr]v(\alpha)
      -v(n_\kappa).
\end{split}
    \label{eq:chainval}
\end{equation}
The valuation bound of~\cite{SibuyaSperber1981} excludes algebraic fugacities that are not algebraic integers. Polynomial dependence of the coefficients on $z$ then excludes transcendental fugacities by the specialization argument of Sec.~\ref{sec:review}.

It remains to locate the effective model in the physical parameter region for all $K>0$ at fixed $H>0$. The transfer matrix of a single bond is
\begin{equation}
    T=D^{1/2}
      \begin{pmatrix}e^K&e^{-K}\\e^{-K}&e^K\end{pmatrix}
      D^{1/2},\qquad
    D=\mathrm{diag}(e^H,e^{-H}).
    \label{eq:transfer}
\end{equation}
For $K>0$, the middle matrix has eigenvalues $e^K\pm e^{-K}>0$. Hence $T$ is positive definite, with eigenvalues $\mu_+>\mu_->0$. Removing the endpoint fields from $T^\kappa$ gives the unnormalized chain weights
\begin{equation}
    B_{st}=e^{-H(s+t)/2}(T^\kappa)_{st}
          =e^{\kappa K+(\kappa-1)H}a_{st},
    \label{eq:chainB}
\end{equation}
and
\begin{equation}
    U_\kappa=\frac{B_{+-}^2}{B_{++}B_{--}},\qquad
    Z_\kappa=z\Bigl(\frac{B_{--}}{B_{++}}\Bigr)^2.
    \label{eq:chainUZB}
\end{equation}
With $E=\mathrm{diag}(e^{-H/2},e^{H/2})$, we have $B=ET^\kappa E$ and $\det E=1$. Therefore,
\begin{equation}
    B_{++}B_{--}-B_{+-}^2
    =\det T^\kappa=(\mu_+\mu_-)^\kappa>0.
    \label{eq:chaindet}
\end{equation}
All chain weights are positive, so $0<U_\kappa<1$: eliminating the internal spins preserves ferromagnetic coupling.

The sign of the effective field follows from the relative weights of the two aligned endpoint configurations. By the Cayley--Hamilton theorem,
\begin{equation}
\begin{split}
    T^\kappa&=xT+yI,\\
    y&=\frac{\mu_+\mu_-
       (\mu_-^{\kappa-1}-\mu_+^{\kappa-1})}
       {\mu_+-\mu_-}<0
\end{split}
    \label{eq:cayley}
\end{equation}
for $\kappa\ge2$. Consequently,
\begin{equation}
    B_{++}=xe^K+ye^{-H},\qquad
    B_{--}=xe^K+ye^H,
    \label{eq:Bpm}
\end{equation}
and $B_{++}-B_{--}=-2y\sinh H>0$ when $H>0$. Thus $0<Z_\kappa<z$, so the effective field remains positive and is larger than the original field.

The effective parameters are real analytic in $K>0$ and remain in the ferromagnetic region with positive field. Joint analyticity of the square-lattice pressure~\cite{Ott2020}, together with~\eqref{eq:chainPhi}, therefore gives analyticity of $p_\kappa$ along the entire physical curve. As for the Lieb lattice, an algebraic differential relation on any open coupling interval would extend to the low-temperature region and contradict the coefficient growth. Spin reversal covers $H<0$, completing the proof of Theorem~\ref{thm:decorated}.

\section{Conclusion}
\label{sec:conclusion}
We have shown in Theorem~\ref{thm:square} that the pressure and the magnetization of the square-lattice Ising model obey no algebraic differential equation on any interval of the coupling at every fixed nonzero real field. Together with~\cite{LuoTian2026}, we thereby rule out solutions of Onsager's type and closed forms in elliptic, hypergeometric, or Painlev\'e functions with fixed parameters at every nonzero field, with no exceptional fields left on the square lattice. The key is the zero-free large-order asymptotic of Proposition~\ref{prop:asym} for the full pressure coefficients at every fugacity outside the unit circle, $L_{2r}(b)=b^{r^2}[\cA(1/b)+o(1)]$ for $|b|>1$, with the amplitude~\eqref{eq:product}. The asymptotics survive the elimination of local spins, so by Theorem~\ref{thm:decorated} the free energies of the Lieb lattice and of the other chain-decorated square lattices with uniform coupling and field obey no algebraic differential equation at any nonzero real field either, although in zero field they follow from Onsager's solution. In every dimension the obstruction also holds at every algebraic fugacity with a real conjugate larger than one, by Theorem~\ref{thm:real}, and for coupled square layers it holds at each fixed field for all but countably many interlayer couplings, by Corollary~\ref{cor:layers}.

In all cases we use the competition between the area and the boundary of droplets, $|S|\le A_{m(S)}$ on the square lattice and~\eqref{eq:isoperimetric} on $\Z^d$. At a fugacity that is not an algebraic integer, the square droplets show up as denominators of the low-temperature coefficients, $v(L_{2r}(z))=v(z)\,r^2$ for every valuation $v$ with $v(z)<0$. At an algebraic-integer fugacity they leave no trace in the denominators or in the size of the physical coefficients and appear only after Galois conjugation, at a conjugate fugacity $\beta$ outside the unit circle, where the weight $\beta^{|S|}$ of a droplet $S$ grows with its area. There large droplets dominate, as in the nucleation of a metastable phase~\cite{Langer1967}, and the near-square droplets make the coefficients grow like $|\beta|^{r^2}$. An algebraic differential equation for the equilibrium free energy would also hold for this conjugated series, so the hidden growth of the droplets excludes it. In an $\ell$-adic valuation the square term has the most negative valuation, and since $v(x+y)=\min\{v(x),v(y)\}$ whenever $v(x)\ne v(y)$, it alone fixes the valuation of $L_{2r}$. At a real conjugate, positivity plays this role in every dimension, because single droplets cannot cancel and the terms of several droplets involve only lower orders. At a complex conjugate the droplets could cancel, and the nonvanishing of the theta function in~\eqref{eq:product} prevents this. For a real conjugate the conjugated series is the formal low-temperature expansion about the all-plus state in a field that favors minus spins.

In two dimensions only the roots of unity remain open, since a nonzero algebraic fugacity $z$ escapes both arguments only if every $\ell$-adic absolute value of $z$ is at most one and $|\sigma(z)|\le1$ for every embedding $\sigma$, and by Kronecker's theorem~\cite{Kronecker1857} this holds only for the roots of unity. A root of unity and all its conjugates lie on the circle $|z|=1$ of imaginary fields, which carries the Lee--Yang zeros of the partition function~\cite{LeeYang1952}, and all its $\ell$-adic absolute values equal one. Hence the area of a droplet does not affect the size of its weight $z^{|S|}$ in any absolute value, and the competition between area and boundary on which our arguments rest disappears. The two solved fugacities, $z=1$~\cite{Onsager1944} and $z=-1$~\cite{LeeYang1952}, are the only real roots of unity, and whether the free energy is differentially algebraic at other roots of unity remains open. In three and more dimensions, where all fields outside a countable set were excluded in~\cite{LuoTian2026}, we now also exclude the algebraic integers with a real conjugate larger than one. In a companion work in preparation~\cite{LuoTian2026c} we investigate the remaining algebraic-integer fugacities in $d\ge3$ through near-cube droplets and higher-dimensional analogues of~\eqref{eq:Fstable} and~\eqref{eq:keybound}.

\par\medskip\noindent\textit{Data availability.---}No data were created or analyzed in this study.

\begin{acknowledgments}
R.L. is supported by National Natural Science Foundation of China under Grant No.~12422503. J.T. is supported by National Natural Science Foundation of China under Grant No.~12405085 and by the Natural Science Foundation of Shanghai (Grant No.~24ZR1419300). J.T. would like to thank Ying Zhang for her love and support. The authors acknowledge DeepSeek which is used to polish the draft grammatically.
\end{acknowledgments}

\bibliographystyle{apsrev4-2}
\bibliography{ising2_references}

\end{document}